\pdfoutput=1 

\documentclass[11pt,a4paper]{article}

\usepackage{jinstpub}

\usepackage{subfigure}

\usepackage{subscript}

\usepackage{amsmath}

\usepackage{multicol}

\usepackage[sort&compress]{natbib}

\usepackage{caption}

\usepackage{multirow}

\title{Working principle and demonstrator of microwave-multiplexing for the HOLMES experiment microcalorimeters}

\author[a]{D.T.~Becker} 
\author[a]{D.A.~Bennett}
\author[b,c]{M.~Biasotti}
\author[d,e]{M.~Borghesi}
\author[b,c]{V.~Ceriale}
\author[c]{M.~De Gerone}
\author[d,e]{M.~Faverzani}
\author[d,e]{E.~Ferri}
\author[a]{J.W.~Fowler}
\author[c]{G.~Gallucci}
\author[a]{J.D.~Gard}
\author[d,e,1]{A.~Giachero}{\note{Corresponding author.}}
\author[c,a]{J.P.~Hays-Wehle}{\note{currently at NASA Goddard Space Flight Center, Greenbelt, MD 20771, USA}}
\author[a]{G.C.~Hilton}
\author[a]{J.A.B~Mates}
\author[d,e]{A.~Nucciotti}
\author[d,e]{A.~Orlando}{\note {currently at Cardiff University, Cardiff CF24 3AA, UK}}
\author[e]{G.~Pessina}
\author[d,f]{A.~Puiu}
\author[a]{C.D.~Reintsema}
\author[a]{D.R.~Schmidt}
\author[a]{D.S.~Swetz}
\author[a]{J.N.~Ullom}
\author[a]{L.R~Vale}

\affiliation[a]{National Institute of Standards and Technology, Boulder, CO 80305, USA}
\affiliation[b]{Dipartimento di Fisica, Universit\`{a} di Genova, Genova I-16146 - Italy}
\affiliation[c]{INFN - Sezione di Genova, Genova I-16146 - Italy}
\affiliation[d]{Dipartimento di Fisica, Universit\`{a} di Milano-Bicocca, Milano I-20126 - Italy}
\affiliation[e]{INFN - Sezione di Milano Bicocca, Milano I-20126 - Italy}

\emailAdd{Andrea.Giachero@mib.infn.it}

\abstract{The determination of the neutrino mass is an open issue in modern particle physics and astrophysics. The direct mass measurement is the only theory-unrelated experimental tool capable to probe such quantity. The HOLMES experiment aims to measure the end-point energy of the electron capture (EC) decay of \textsuperscript{163}Ho with a statistical sensitivity on the neutrino mass as low as $\sim 1$\,eV/c$^2$. In order to acquire the large needed statistics, by keeping the pile-up contribution as low as possible, 1024 transition edge sensors (TESs) with high energy and time resolutions will be employed. Microcalorimeter and bolometer arrays based on transition edge sensor with thousands of pixels are under development for several space-based and ground-based applications,  including astrophysics, nuclear and particle physics, and materials science. The common necessary challenge is to develop pratical multiplexing techniques in order to simplify the cryogenics and readout systems. Despite the various multiplexing variants which are being developed have been successful, new approaches are needed to enable scaling to larger pixel counts and faster sensors, as requested for HOLMES, reducing also the cost and complexity of readout. A very novel technique that meets all of these requirements is based on superconducting microwave resonators coupled to radio-frequency Superconducting Quantum Interference Devices, in which the the changes in the TES input current is tranduced to a change in phase of a microwave signal. In this work we introduce the basics of this technique, the design and development of the first two-channel read out system and its performances with the first TES detectors specifically designed for HOLMES. In the last part we explain how to extend this approach scaling to 1024 pixels.}

\begin{document}

\maketitle
\flushbottom

\section{Introduction}
The observation of neutrino oscillations is a clear evidence that neutrino are massive particles. However, as neutrino oscillation experiments allow only to measure the difference between neutrino mass states~\cite{Capozzi2017,deSalas2017}, the assessment of neutrinos absolute mass scale is still an outstanding challenge~\cite{Capozzi2018}. The kinematic analysis of electrons emitted in single $\beta$ or electron capture (EC) decay is the only theory-unrelated experimental approach to effectively determine the electron-neutrino mass~\cite{Weinheimer2013,Drexlin2013}. The method consists in searching for a tiny deformation caused by a non-zero neutrino mass in the spectrum near its end point. The most stringent results on $m_{\nu_e}$ come from electrostatic spectrometers used to study the Tritium $\beta$ decay, setting an upper limit on the neutrino mass of $m_{\nu_e} < 2.05$\,eV/$c^2$ at 95\%\,C.L.~\cite{Mainz,Mainz2,Troitsk,Troitsk2}. The current generation of tritium end point experiment based on same detector technology is the Karlsruhe Tritium Neutrino (KATRIN) experiment~\cite{KATRIN}. KATRIN is a very large scale tritium electrostatic spectrometer that aims to deliver an order-of-magnitude improvement in neutrino mass sensitivity: 0.2~eV/$c^2$ in five years~\cite{KATRIN2017} of data taking. The experiment reaches the maximum size and complexity practically achievable for a spectrometer experiment, and no further improved project can be presently envisaged. A new approach is required to go beyond the 0.2\,eV limit. Project 8~\cite{Project8,Project8_3} is a phased concept for a next-generation tritium beta spectrometry experiment based on the detection of the relativistic cyclotron radiation emitted by the beta electrons when trapped in a magnetic field~\cite{Project8_2}. An alternative experimental approach  to spectrometry and with the potential to reach the sub-eV sensitivity is provided by calorimetric measurements. The beta or EC source is embedded in the detector and the energy emitted in the decay is entirely measured by the detector, except for the fraction taken away by the neutrino. This approach eliminates both the problematic of an external source and the systematic uncertainties coming from the excited final states. Phonon-mediated detectors operated at low temperatures are particularly suitable for such measurements~\cite{UseOfLTD}.

Although the calorimetric measurement of the energy released in the EC decay of \textsuperscript{163}Ho was proposed in 1982 by A. Rujula and M. Lusignoli~\cite{HoProposal} as an appealing method for directly measuring the electron neutrino mass, only in the last decade the technological progress in detector development and multiplexing allowed to design competitive experiments capable of reaching a sensitivity as low as 1\,eV on the neutrino mass. Besides the HOLMES~\cite{HOLMES} experiment, which is the main topic of this contribution, there are currently two other experiments aiming to exploit the \textsuperscript{163}Ho electron-capture spectroscopy for measuring the neutrino mass. ECHo~\cite{ECHo} uses arrays of Metallic Magnetic Calorimeters~\cite{MMC_REVIEW} with \textsuperscript{163}Ho ion implanted. In the US the NuMECS collaboration is critically assessing the various technologies required to produce, separate, and embed the \textsuperscript{163}Ho isotope~\cite{HOLMES_Ho,Engle2013}. NuMECS plans to measure the \textsuperscript{163}Ho with arrays of transition edge calorimeters~\cite{NuMECS}. For all the three experiments a very crucial challenge for reaching the desired sensitivity~\cite{Nucciotti_sens} is to operate a large array of detectors in order to collect a high amount of events at the end-point of the spectrum. This calls for the implementation of an efficient multiplexing system for reading out this large amount of detectors with the smallest possible number of amplifiers and read out lines. 

The neutrino mass sensitivity is strongly dependent on the unresolved pile-up fraction in the end point region. This happens when two events of energies $E_1$ and $E_2$ occurs within a time interval shorter than the time resolution of the detector giving a resulting event acquired and processed as a single event with an energy $E \simeq E_1 + E_2$. The better is the detector time resolution the better is the pile-up recognition capability. The multiplexing system must provide a read out bandwidth compatible with the necessary time resolution. The read out bandwidth remains one of the most serious technical constraint to the full exploitation of low temperature calorimeters for the direct measurement of the neutrino mass.

\section{TES Sensors and Multiplexing techniques}
A Transition Edge Sensor (TES) exploits the resistive phase transition of a superconductor to obtain a thermometer~\cite{TES_MODEL,TES_REVIEW}. Its logarithmic sensitivity $\alpha =d \log{R} / d\log{T}$ can be more than an order of magnitude higher than for a practical semiconductor thermistor. TES-based micro-calorimeters can be read out monitoring the sensor resistance, since the resistance variation $\Delta R$ is a function of the radiation energy deposited in the micro-calorimeter absorber. The monitoring is generally done by applying a constant voltage across the TES resistance, and reading the temperature-dependent current. A voltage-biased TES is stabilized by strong electrothermal feedback (ETF)~\cite{ETF}. In fact by keeping constant the bias voltage, an increase in temperature in the absorber yields a sharp increase in resistance, which reduces the current flowing through the TES, lowering the dissipated power and decreasing the temperature. This enables the devices to be very fast and has the additional benefit of reducing Johnson noise so that phonon noise dominates at low frequencies. 

For typical TES operating resistance $R_0 < 10\,\mbox{m}\Omega$, the current signal is of the order of few microamps with noise fluctuations around (100-200)\,pA/$\sqrt{\mbox{Hz}}$. Non-degrading measurement of such small and quiet currents requires a low noise, low power, and low impedance amplifier. These requirements rule out the use of classical semiconductor amplifiers and leave Superconducting Quantum Interference Device (SQUIDs) amplifiers as the only viable option. A SQUID consist of a superconducting loop interrupted by one or more Josephson junctions~\cite{SQUID}. The  direct current (dc) SQUID~\cite{DCSQUID} consists of a superconducting loop interrupted by two resistively-shunted Josephson junctions, while in the radio frequency (rf) SQUID~\cite{RFSQUID} the loop is interrupted by a single resistively-shunted Josephson junction. These devices are magnetometers sensitive to magnetic flux that is a fraction of the superconducting magnetic flux quantum $\Phi_0 \simeq 10^{-15}$\,Wb. SQUIDs typically have flux noise of few $\mu\Phi_0/\sqrt{\mbox{Hz}}$ that is sufficient to give a current noise of around (20-40)\,pA/$\sqrt{\mbox{Hz}}$, well below the current noise of many TES typologies.

TESs are typically read out by dc-SQUID. When a dc-SQUID is biased at a current below the junction critical current $I_c$\footnote{The maximum supercurrent (current flowing with no voltage) that can flow through a Josephson junction.}, when an external variable magnetic flux is coupled into the Josephson loop the voltage drop across the Josephson junction changes. Monitoring the change in voltage allows to determine the magnetic flux variation $\Delta\Phi$ applied to the SQUID. In this configuration, though, the response of the SQUID is degenerate: its output is periodical in the magnetic flux (e.g. 1$\Phi_0$ produces the same voltage output as 2$\Phi_0$ and as 3$\Phi_0$). To avoid this, dc-SQUIDs are operated in a flux-locked loop, in which a feedback flux is applied to the SQUID to cancel the change in flux from the input coil. This "locks" the SQUID in a state of constant flux and the change in the feedback current, coupled into the SQUID loop, is then a measure of the external flux variation. A voltage-biased TES in series with a pickup inductor coupled with the SQUID induces a change in the flux through the SQUID when the TES resistance changes.   

\begin{figure}[!t]
\centering 
\includegraphics[width=\textwidth,clip]{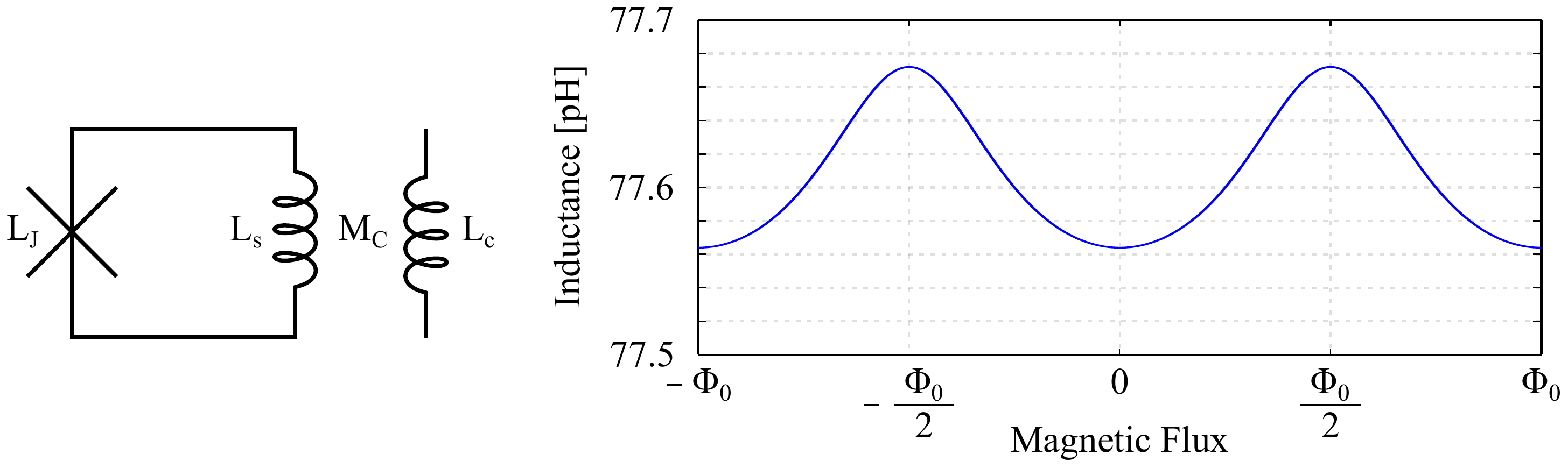}
\caption{\label{fig:squid-response}(left) Circuit diagram of an rf-SQUID screening another inductor. (right) Effective flux-variable inductance for an non-hysteretic rf-SQUID. Plot edited from~\cite{MATES_THESIS}.}
\end{figure}

In recent times also rf-SQUID have been used to read out TESs detectors. A dissipationless rf-SQUID consists of a superconducting loop closed by a single unshunted Josephson junction with Josephson inductance $L_J$, loop inductance $L_s$ and critical current $I_c$. For $\beta_L=2\pi L_sI_c/\Phi_0<1$ (small flux oscillations condition) it behaves purely reactive impedance, and can be modelled as a non-linear inductor whose value depends on the magnetic flux external to the SQUID. In order to read out the change of impedance caused by a magnetic flux variation associated with a detector signal, the rf-SQUID is inductively coupled to a load inductor $L_c$ (figure \ref{fig:squid-response}, left) part of a high quality factor superconducting microresonator with resonance frequency in the GHz range. Due to the mutual interaction with the rf-SQUID, the value of the equivalent inductor $L(\Phi)$ is also dependent on the flux (figure \ref{fig:squid-response}, right) according to~\cite{MATES_THESIS}

\begin{equation}\label{eq:var-imp}
L(\Phi)=L_c-\cfrac{{M_c}^2}{L_s}
\left( 
\cfrac{\lambda \cos{\left(2\pi\,\Phi/\Phi_0\right)}}{1+\cos{\left(2\pi\,\Phi/\Phi_0\right)}}
\right)
\end{equation} 

\noindent where $\lambda=L_s/L_J$ and $M_c$ is the mutual inductance between the rf-SQUID and the load inductor. The microresonator resonance frequency shifts as the inductance $L(\Phi)$ changes. Since $L(\Phi)$ depends on the magnetic input flux, by monitoring the variations in the resonance parameters it is possible to evaluate the TES resistance changes~\cite{MSQUID}.

Multiplexed readout implies that a group of TES detectors are readout by a single amplification line, and voltage biased by one wire pair. Multiplexed readout as opposed to single (or direct) readout  is mainly driven by the goal to minimize the required cooling capacity and complexity at the cold stages. Without multiplexing, a large amount of wire pairs are required for operating the dc-SQUID amplifiers, i.e. around 4k pairs for the HOLMES detector. By implementing a multiplexing technique it is possible to reduce the number of wires resulting in the reduction of volume, system complexity and heat transfer to the cryogenic stage.

After more than ten years of development, multiplexing techniques for TESs has been demonstrated~\cite{Shannon} in the time~\cite{TDM}, frequency~\cite{FDM} and code~\cite{CDM} domains (TDM, FDM and CDM). The most mature multiplexing technology at present is time-division multiplexing that is deployed in many astronomical instruments~\cite{Doriese2016} and beamline and laboratory science~\cite{beam}. In TDM the TES outputs are switched by applying bias current to one SQUID amplifier at a time. A two-dimensional (M rows $\times$ N columns) array of pixels can be read out by sequentially turning on the SQUIDs in every column, one row at a time. The typical multiplexing factor is around 30 . The basic limit on this is maximum total read out rate that can not exceed more than 6.25\,MSamples/second~\cite{Doriese2016,beam}, value that has to be shared among all 30, in a column. This means a sampling rate per channel of around 200\,kHz (sampling time around 5\,$\mu$s), with a Nyquist bandwidth of around 100\,kHz. In these conditions it is possible to multiplex TESs with pulses with rise and fall exponential time constants of around 60 and 1000\,$\mu$s~\cite{Doriese2016}, respectively. 

In Frequency Domain Multiplexing (FDM) each TES signal is amplitude modulated by a sinusoidal bias current of a distinct frequency per pixel. The current through the TES is coupled to a SQUID and all the SQUID outputs are summed on a single line by using a common transformer coil~\cite{FDM}. The frequency separation between pixels is driven by cross talk requirements, and SQUID dynamic range. A 100 kHz separation between adjacent carrier frequencies fulfills these requirements and is compatible with a bandwidth per pixel of around 10 - 15 kHz. FDM can provide a larger multiplexing factor ($>40$) over a total bandwidth of 5\,MHz~\cite{ATHENA_MUX}. FDM is currently the baseline readout system for the Athena X-Ray Integral Field Unit (X-IFU)~\cite{ATHENA}

For both TDM and FDM, the only possibility to acquire faster pulses with the same total bandwidth is to decrease the multiplexing factor. Future applications, as neutrino mass measurement and spectrometer at free-electron laser facilities, need pulse times below 200\,$\mu$s with a very fast time resolution in order to facilitate the pile-up rejection due to the repetition rates of the source. A significantly faster pulse response (higher bandwidth), higher multiplexing factor ($>1000$), and lower power dissipation are fundamental requirements for future experiments. Standard multiplexing technologies, like TDM are reaching their full potential and can not match these requirements. A multiplexing technique capable of reading out thousands of detectors in a single output channel needs a total bandwidth of a few gigahertz. Microwave SQUID multiplexer of dissipationless rf-SQUIDs is a promising techniques that can meet these requirements.  

\begin{figure}[!t] 
\centering 
\includegraphics[width=\textwidth,clip]{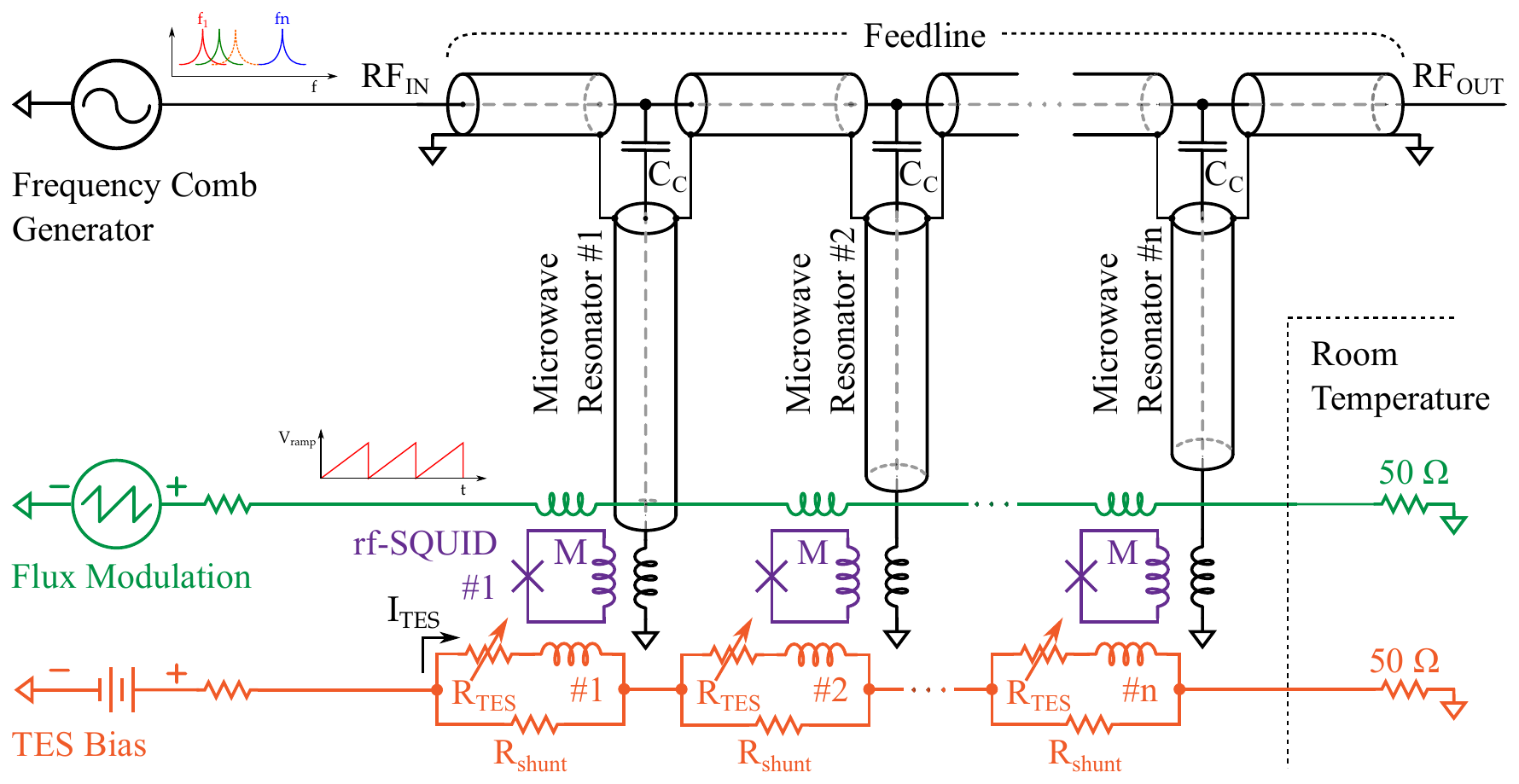}
\caption{\label{fig:mw-squid} Basic elements of microwave SQUID multiplexed TES readout: (black) microwave multiplexer or $\mu$mux, (green) flux-ramp demodulation line, (violet) rf-SQUID, (orange) TESs and theirs DC-bias line.}
\end{figure}

\section{Microwave SQUID multiplexing}\label{sec:umux}
The basic concept of the microwave SQUID multiplexing applied to TES detectors was proposed for the first time by Irwin et al.~\cite{IrwinMSQUID} and Hahn et al.~\cite{HahnMSQUID}. The technique is inspired by the GHz frequency domain multiplexing approach of Microwave Kinetic Inductance Detectors (MKIDs)~\cite{DayMKIDS,ZmuidzinasMKIDS}.

As reported in~\cite{MSQUID}, each channel consists of a dc-biased TES (orange in figure \ref{fig:mw-squid}) inductively coupled to a dissipationless rf-SQUID (violet in figure \ref{fig:mw-squid}), inductively coupled to a high-Q superconducting quarter-wave resonator (black in figure \ref{fig:mw-squid}). The input channels are coupled to a common microwave read out line (feedline) through capacitive coupling ($C_c$). 
When probed with a sinusoidal signal matched to its resonant frequency, the resonator acts as a short to ground (notch filter). 

The energy deposition in the microcalorimeter absorber is detected by changes in the TES resistance, that modulates the $I_{\scalebox{.6}{\mbox{TES}}}$ current. This time-dependent current applies a flux $\Phi$ in the SQUID loop as $\Phi=M I_{\scalebox{.6}{\mbox{TES}}}$, where $M$ is the SQUID transduction gain. Since the rf-SQUID acts as a flux-variable inductor (equation \ref{eq:var-imp}), the shift in resonant frequency causes a change in amplitude and phase of the microwave sinusoidal signals in the feedline coupled to the resonator. This change corresponds to an increase of the power transmitted through the feedline for the corresponding probe tone. By placing many micoresonators with different resonant frequencies in the same multiplexer chip it is possible to perform the read out of multiple detectors. The detector array can be monitored by a set of sinusoidal probe tones (frequency comb). A cryogenic High Electron-Mobility Transistor (HEMT), placed at cryogenic temperatures ($\sim 4$\,K), amplifies the transmitted microwave signals at the multiplexer output. 

\begin{figure}[!t]
\centering 
\includegraphics[width=\textwidth,clip]{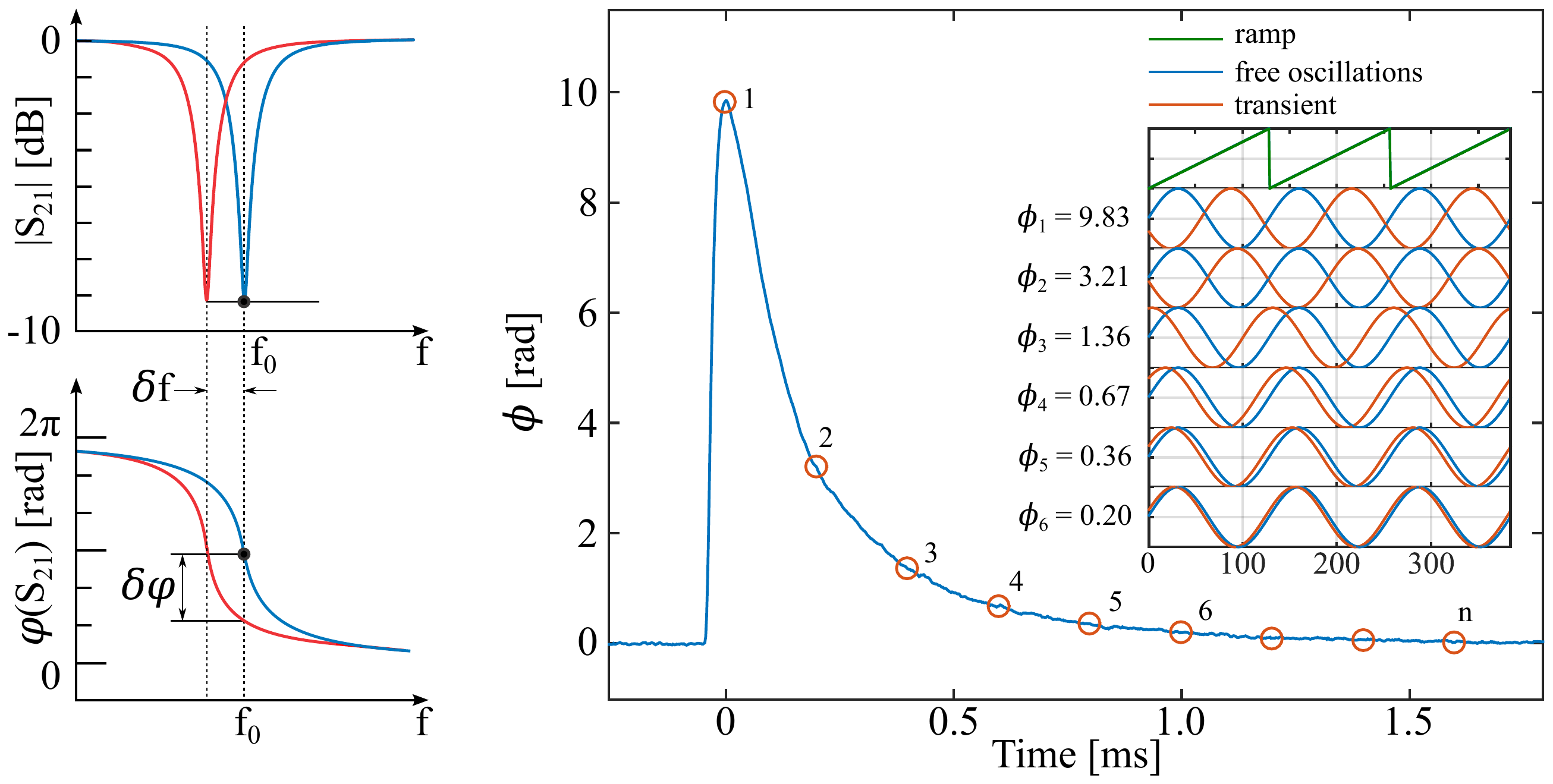}
\caption{\label{fig:shift-dem} (left) Frequency and phase shift in response to a change in the rf-SQUID input flux. $S_{21}$ is the forward S-parameter of the resonator. (right) Ramp-demodulated signal and related with phase shifts in the rf-SQUID response (inlet). In this example each change in the oscillation phase (i.e. signal sample) is demodulated via a ramp amplitude tuned to provide $n_{\Phi_0}$ of flux per ramp.}
\end{figure}


SQUIDs are interferometers with a periodic and therefore nonlinear response (i.e. figure \ref{fig:squid-response}, right), which are traditionally linearized by a flux-locked loop. This approach is not compatible with the microwave SQUID multiplexer since a flux-locked loop would require a separate feedback wire for each SQUID, which is in contrast with the multiplexing purpose. An alternative way is to implement a \textit{flux-ramp modulation} scheme that modulates the signal from a rf-SQUID and linearizes the response of all pixels using a single wire~\cite{RAMP,RAMP2}. In this scheme, a common flux ramp is applied to all SQUIDs. The flux ramp is a sawtooth signal with a frequency of $f_{\scalebox{0.6}{\mbox{ramp}}}$ and with an amplitude of an integer number $n_{\Phi_0}$ of flux quanta, that induce a periodic oscillation in the SQUID response. The detector signal is measured as a change in the phase of the periodic SQUID response. If the slew-rate of the applied ramp exceeds that of any input signal, a variation in the input signal looks like a flux offset during the ramp period, which produces a phase shift in the SQUID response respect to the free oscillation state, i.e SQUID oscillation in absence of TES signal, (figure \ref{fig:shift-dem}, right). This phase shift within the  i\textsuperscript{th}-ramp period is given by  
\begin{equation}\label{eq:phi-angle}
\phi_i=2\pi\cfrac{\Phi_i}{\Phi_0}
\end{equation}
\noindent where $\Phi$ is input flux variation due to the TES. This is a linear function of the detector signal and can be tracked through many flux quanta. 



Considering a purely sinusoidal SQUID response $\theta(t)=\cos{\left(\omega_ct\right)}$, where $\omega_c=2\pi\,n_{\Phi_0}f_{\scalebox{0.6}{\mbox{ramp}}}$ is the angular frequency for the SQUID free oscillations ($\phi_i=0$), a flux variation $\Phi_i$ produced by any changes in the input signal within the ramp period can be written as $\theta(t,\phi_i)=\cos{\left(\omega_ct+\phi_i\right)}$. As suggested in~\cite{RAMP,MATES_THESIS}, a simple Fourier measurement can estimate the phase angle $\phi_i$ of the fundamental frequency in the SQUID response. Introducing a carrier frequency equal to the free-oscillation response frequency $\omega_c$ and multiplying the SQUID response by sine and cosine at that frequency, the phase angle $\phi_i$, during the duration of the ramp, can be extracted as
\begin{equation}\label{eq:demod-phi}
\phi=\arctan{
\left[
- \cfrac
{\sum\limits\theta(t,\phi) \sin{\left(\omega_ct\right)}}
{\sum\limits\theta(t,\phi) \cos{\left(\omega_ct\right)}}
\right]
}
\end{equation}

\noindent This solution is computationally efficient and the knowledge of the precise response function of the SQUID is not required. The assumption of a purely sinusoidal response for the SQUID has a little impact, since higher harmonics affect the read out noise at negligible level thanks to matching couplings. Therefore the SQUID response can be considered purely sinusoidal~\cite{RAMP,MATES_THESIS}. More details on how to derive the relationship \ref{eq:demod-phi} are reported in Appendix \ref{sec:appendix-flux-ramp}. The measurement of the SQUID response $\theta(t,\phi_i)$ can be performed by monitoring the corresponding resonator. In fact, the change in the phase angle $\phi_i$ is transduced in amplitude and phase variations of the microwave signal transmitted through the feedline.

\begin{figure}[!t]
\centering 
\includegraphics[width=\textwidth,clip]{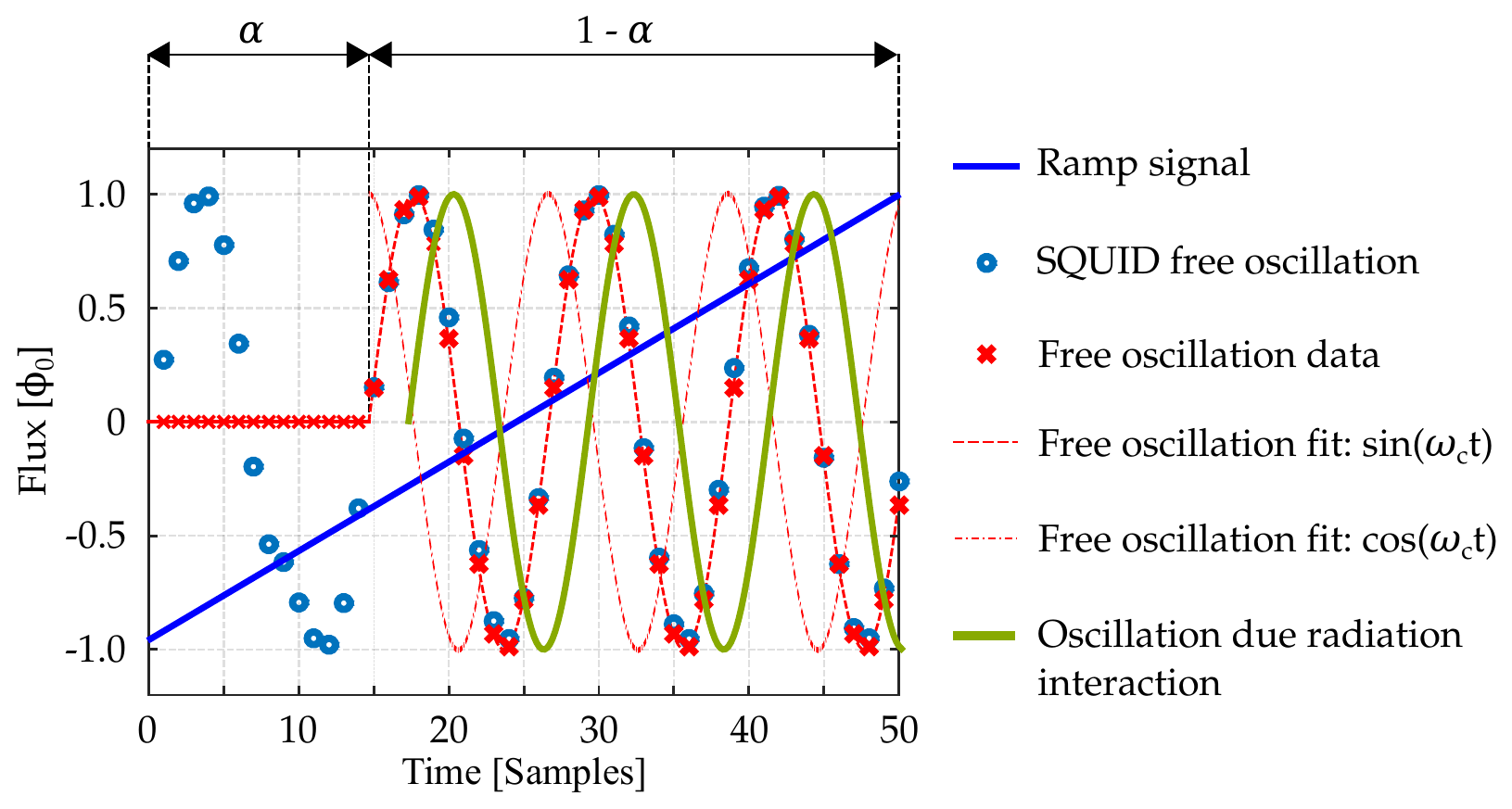}
\caption{\label{fig:reset} The phase angle $\phi$ is the arctangent of the ratio actual data of a ramp response (green) multiplied by both red curves. The first SQUID oscillation during the ramp period is discarded to let the ramp reset transient settle. $\alpha$ is the discarded fraction of the total ramp.}
\end{figure}

To prevent contamination in the demodulated data from unwanted transient behavior caused by the ramp reset, a fraction of oscillations at the beginning of the ramp period is discarded. This is implemented by multiplying the demodulation carriers, $\sin{\left(\omega_ct\right)}$ and  $\cos{\left(\omega_ct\right)}$, times a Heaviside step function $\Theta(t-T_{\scalebox{.6}{\mbox{ramp}}}/\alpha)$, where $T_{\scalebox{.6}{\mbox{ramp}}}$ is the ramp period and $\alpha$ is the discarded fraction of the ramp within the total period (figure \ref{fig:reset}). For $t<T_{\scalebox{.6}{\mbox{ramp}}}/\alpha$ the sum contributions in the relationship \ref{eq:demod-phi} is zero and only the samples with $t>T_{\scalebox{.6}{\mbox{ramp}}}/\alpha$ contribute at the ramp demodulation. In case of $n_{\Phi_0}=2$ flux oscillations per ramp, that is the minimum achievable number, $\alpha$ has to be at most 0.5.

\begin{figure}[!t]
\centering 
\includegraphics[width=0.85\textwidth,clip]{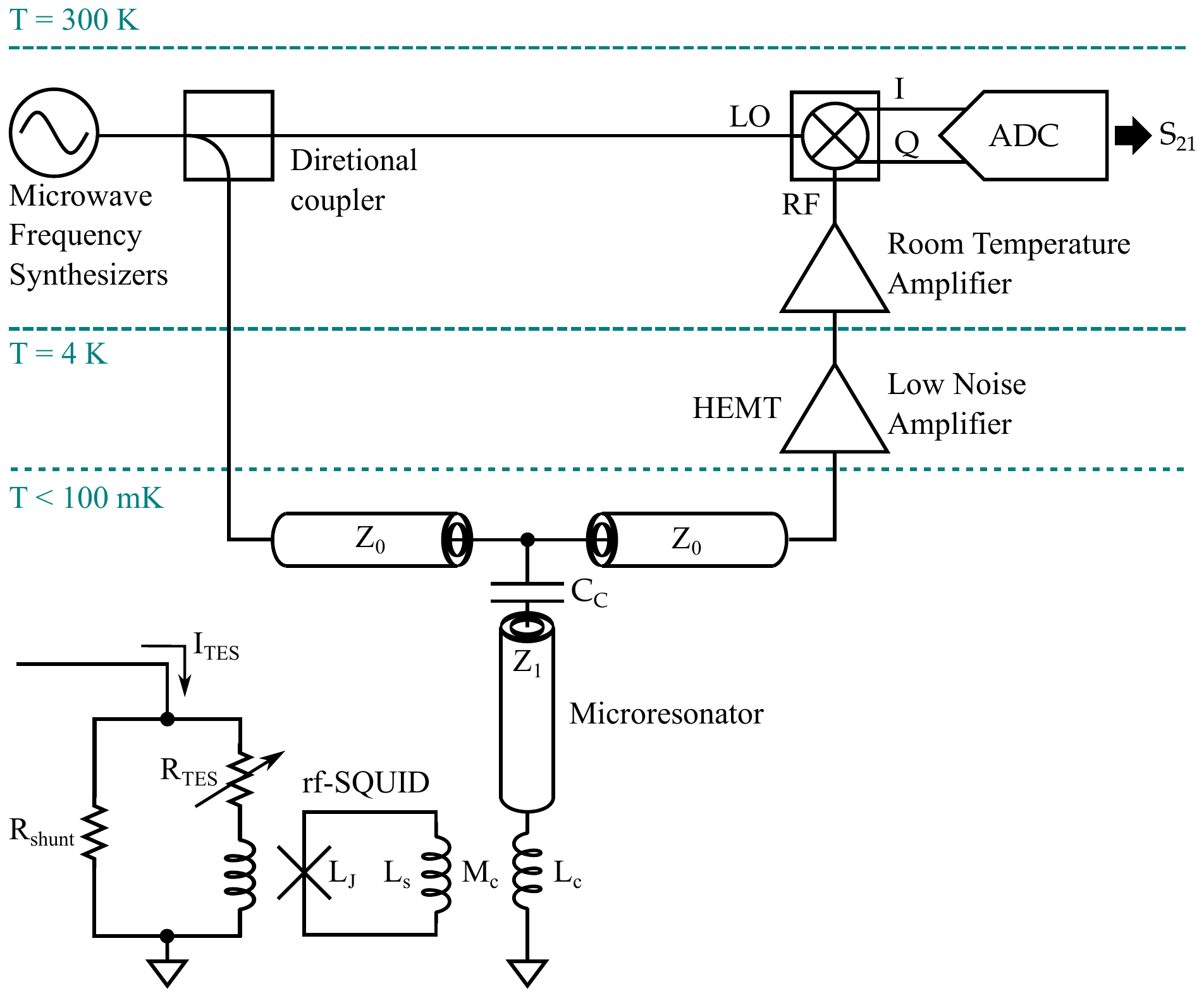}
\caption{\label{fig:homo} The homodyne detection scheme employed to read out rf-SQUIDs. $L_J$ and $L_S$ are the rf-SQUID Josephson and loop self inductance, respectively, $L_C$ the micro-resonator coupling inductance and $M_C$ the mutual inductance between the rf-SQUID and the micro-resonator.}
\end{figure}

\begin{figure}[!t]
\centering 
\includegraphics[width=\textwidth,clip]{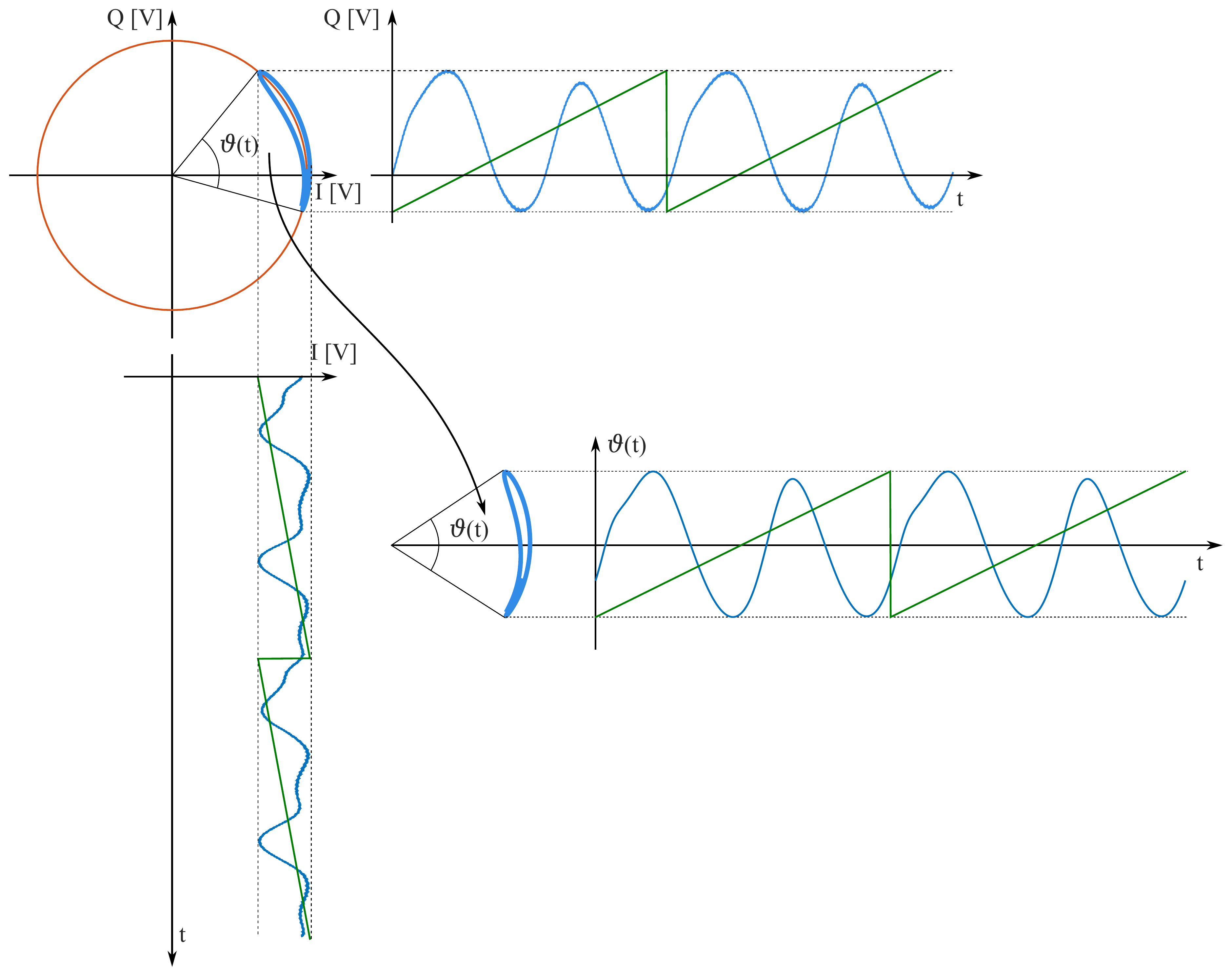}
\caption{\label{fig:IQcircle} In-phase $I$ and quadrature $Q$ oscillations on the IQ-plane generated by the ramp modulation. Real data from Run \#26}
\end{figure}

A possible method to monitor the resonators response is the \textit{homodyne detection}, which is essentially a dual-phase lock-in detection technique. A microwave signal at the resonator resonant frequency is generated by a synthesizer and then split in two (figure \ref{fig:homo}). One copy is sent into the cryostat and then into the transmission line of the multiplexer chip, through a HEMT amplifier, and back out at room temperature to an IQ mixer (RF), where it is mixed with the second copy of the original microwave signal (LO). The IQ mixer actually contains a quadrature coupler (a 90-degree hybrid coupler), a splitter, and two mixers so that a copy of the incoming signal is mixed with the original microwave signal (in-phase signal $I$) and another copy of the incoming signal is mixed with a 90-degree phase-shifted copy of the synthesized signal (quadrature signal $Q$). The outputs $I$ and $Q$ represent the real and imaginary parts of the forward-transmission parameter $S_{21}$~\cite{Pozar}. The periodic variations in amplitude and phase of $S_{21}$ produce periodic oscillations on $I$ and $Q$ over a circle in the IQ-plane (figure \ref{fig:IQcircle}). By measuring the signals $I(t)$ and $Q(t)$ (figure \ref{fig:IQcircle}) it is possible to recover the signal $\theta(t,\phi_i)$ as 

\begin{equation}\label{eq:demod-theta}
\theta(t,\phi_i)=\arctan{\left[\cfrac{I(t,\phi_i)-I_c}{Q(t,\phi_i)-Q_c}\right]} 
\end{equation}

\noindent where $(Q_c,I_c)$ is the center of the circle. Then the phase angle $\phi_i$ is obtained with the relation \ref{eq:demod-phi}. The analog homodyne read out system can be easily extended to 2 multiplexed channels by using two different synthesizers, but a further extension is pratical to implement. A key enabling technology for large-scale microwave multiplexing is the digital approach that replaces most of the analog components with digital processing. This exploits standard Software-defined Radio (SDR) techniques used in microwave-frequency communications (more details in Section \ref{sec:1000pixels}). The number of channels that can be read out in parallel with a digital system is theoretically limited by the bandwidth of the DACs and ADCs used to generate the probe tones and to acquire the detectors response (more details in Section \ref{sec:factor}).  

Compared to the other multiplexing techniques, in microwave multiplexing all the electronics is at room temperature except for the HEMT amplifier which is placed at 4\,K. The complexity is moved at the warm stage outside the cryostat, where it is possible to leverage high-bandwidth commercial components and implement many functions using SDR techniques. Microwave SQUID multiplexing has been proposed and it is currently in development also to read out large arrays of Metallic Magnetic micro-Calorimeters (MMC)~\cite{MMC}.  

\section{Noise considerations}\label{sec:noise}
The microwave multiplexing must operate without degrading the detector sensitivity, this means that its noise contribution must be smaller than the intrinsic noise of the sensors under study. As reported in \cite{MATES_THESIS}, the main sources of noise for the microwave multiplexing based on rf-SQUID are: Johnson noise in the flux input circuit, intrinsic flux noise in the SQUID, HEMT noise, and Two-Level System (TLS) noise in the resonator.

\underline{Johnson contribution}: the circuit that couples inductively the microresonator to the SQUID may have an impedance with a real component ($R_f$). Microwave power in the resonator may therefore dissipate in the input circuit. The power dissipation can induce Johnson noise currents through the SQUID input coil with a power density $S_I = 4k_BT/R_f$. Considering a typical resistance around $R_f\simeq 0.2$\,m$\Omega$ and a mutual inductance $M\simeq 88$\,pH, this produces a flux noise around 0.2\,$\mu\Phi_0/\sqrt{\mbox{Hz}}$, which is small compared to the other sources of noise in the system.

\underline{SQUID contribution}: in a dissipationless rf-SQUID there a two different sources. The noise can arise from fluctuations in the critical current across the junction, and from the flipping of magnetic dipoles on the SQUID loop. In either cases the spectral density scales as $1/f$. Measurements with dc-SQUIDs~\cite{TES_REVIEW,MATES_THESIS} have showed that the sum of the two contributions dominates below 1\,$\mu\Phi_0/\sqrt{\mbox{Hz}}$ at 1\,Hz, given a negligible contribution in a TES typical signal power spectrum. Since the fabrication process for rf-SQUID has remained unchanged, it is reasonable to assume that neither of these two noise sources represent a threat.

\underline{HEMT contribution}: the microwave signal passing through the resonators is amplified by a HEMT microwave amplifier. In order to keep the amplifier noise as low as possible, the HEMT is placed at 4\,K. Despite that the cold amplifier noise is by far the most significant component that affects the SQUID readout noise. As reported in \cite{MATES_THESIS}, its noise power spectrum referred to the flux output of the SQUID is given by

\begin{equation}
  S|_{f=f_{\scalebox{.6}{\mbox{pixel}}}\,,\,\Phi=\Phi_0/2}=\cfrac{4k_BT_NL_j}{\pi f_{\scalebox{0.6}{res}}}
\end{equation}

\noindent where $T_N$ is the amplifier noise temperature, $f_{\scalebox{.6}{\mbox{res}}}$ the resonator resonant frequency, $L_j$ the junction impedance, and $k_B$ Boltzmann constant. For $T_N=6$\,K at a readout frequency $f_{\scalebox{.6}{\mbox{res}}}=6$\,GHz and $L_J$=60\,pH, the HEMT amplifier noise is 0.6\,$\mu\Phi_0/\sqrt{\mbox{Hz}}$.  

\underline{Two Level System}: superconducting microwave microresonators are effected by the two-level systems noise~\cite{TLS1,TLS2,TLS_EMPIRICAL} which originates from interactions with dielectric layers on the surfaces~\cite{TLS4}. The fluctuation of these two-level systems introduce a power spectral density that varies with frequency as $f^{-1/2}$, caused by the coupling of the TLS electric dipole moments to the electric field inside the resonator. It was experimentally proved that the TLS noise is generated in the capacitative section of the resonator~\cite{TLS_REDUCTION}. This implies that the TLS noise could be dramatically reduced by decreasing the surface layer to volume ratio of the capacitors, for instance by using interdigitated capacitors (IDC) with large spacing between their fingers. Despite a microscopic theory of TLS noise is not yet available, there is a semi empirical model that describes the contribution of the TLS to the power spectral density of the fractional frequency shift as a function of the driving power $P_g$ and temperature $T$~\cite{TLS_EMPIRICAL}. The spectral density varies as $P_g^{-1/2}$ and $T^{\beta}$ with $\beta=1.5-2$~\cite{Zmuidzinas2012}. Since this noise is dominant at low frequency ($\propto f^{-1/2} $) its contribution can be evaded with a flux-ramp modulation at higher frequencies, ideally where the two-level system noise falls below the HEMT noise~\cite{PUIU_THESIS}.

Finally, the flux-ramp modulation increases readout noise if compared with the flux-locked loop readout, which works in the steepest slope of the SQUID response. The main reasons are two: 
\begin{enumerate}
\item  unlike the flux-locked loop read out, a substantial fraction of each ramp is spent measuring the SQUID at an extremum of its response curve where it is practically unaffected to changes in input flux. Computation performed in \cite{MATES_THESIS} showed that, for purely sinusoidal SQUID response, the flux-ramp modulation imposes a noise penalty of $\sqrt{2}$. 

\item the demodulation is not performed on the full span of the flux ramp, but the reset transient from the beginning of each ramp response is discarded. 
\end{enumerate}

Combining this two contributions, the effective degradation in signal-to-noise ratio for a sinusoidal SQUID response function with flux-independent noise is $\sqrt{2/(1-\alpha})$~\cite{RAMP}, where $\alpha$ is the discarded fraction of the total ramp (as defined in \ref{sec:umux}). This effect is partially compensated by the fact that the ramp modulation evades low-frequency noise added after the SQUID in the amplifier chain (i.e. TLS noise) and only noise components at the carrier frequency affect the measurement of the input signal. 

Recent measurement~\cite{128Mux} showed that with the microwave multiplexing it is possible to achieve a total readout noise level of 2\,$\mu\Phi_0/\sqrt{\mbox{Hz}}$ that means an input-referred current noise added at the TES noise of around 20\,pA$/\sqrt{\mbox{Hz}}$~\cite{128Mux}. Despite this measured valued is larger than the sum of the expected contributions previously discussed, this noise level is low enough to have negligible impact on the resolution of most microcalorimeters, with a typical sensor noise greater than 100\,pA/$\sqrt{\mbox{Hz}}$. The noise performances obtained for the microwave multiplexing demonstrator developed for HOLMES are reported in Section \ref{sec:demo}.

\section{Multiplexing factor}\label{sec:factor}
 Microwave multiplexing is the most suitable read out system for HOLMES, since it provides a larger bandwidth for the same multiplexing factor (number of multiplexed detectors). This novel approach was demonstrated for the first time for gamma-ray spectroscopy~\cite{GAMMA-DEM} and has been proposed for many current and future applications based on superconducting transition-edge sensor where fast pulse response is required~\cite{Argonne,LosAlamos}. 

The base element of the microwave multiplexing is the multiplexer chip, where the microresonators responses are read out through a common feedline. The coupling between a resonator and the feedline determines the bandwidth of the resonance. The resonator bandwidth determines the minimum frequency spacing between adjacent resonances. A good rule to avoid cross-coupling between two adjacent resonators is to have a space between resonances by $(7-10)$ times their bandwidth. The coupling between the resonator and the SQUID determines the shift in resonance frequency in response to magnetic flux in the SQUID. Ideally, peak-to-peak shift in resonance frequency should match the bandwidth of the resonance. To prevent signal roll-off, crosstalk and non-linearity, only a fraction of the available bandwidth can be used. From these considerations, the number of pixel as a function of the available bandwidth is given by:

\begin{equation}\label{eq:n-pixel}
N_{\scalebox{.6}{\mbox{pixel}}}=\cfrac{f_{\scalebox{.6}{\mbox{ADC}}}\cdot\tau_r}{2\cdot n_{\Phi_0}\cdot g_f\cdot R_d}
\end{equation}

\noindent where $f_{\scalebox{.6}{\mbox{ADC}}}$ is the total available bandwidth of the ADC, $\tau_r$ the detector rise rime, $n_{\Phi_0}$ the number of flux quanta per ramp, $g_f$ the guard factor between tones (used to define the space between adjacent resonances by  $S>g_f\Delta f_{\scalebox{.6}{\mbox{BW}}}$, where $f_{\scalebox{.6}{\mbox{BW}}}$ is the resonators bandwidth), and $R_d$ distortion suppression factor (defined as $R_d=\tau_r/t_{\scalebox{.6}{\mbox{ramp}}}$, specifies the distortion level in the demodulation process,  $R_d=2$ is Nyquist limit). More details on how to derive the multiplexing factor are reported in Appendix \ref{sec:appendix-budget}.

Considering a rise time $\tau_r =10\,\mu$s, i.e. the HOLMES target, a distortion suppression factor $R_d = 5$ and the condition $f_r\ge R_d/\tau_r$ (equation \ref{eq:samp}), this means a ramp frequency of $f_{\scalebox{.6}{\mbox{ramp}}} = 500$\,kHz. In case of a number of flux quanta per ramp $n_{\Phi_0}=2$, that is the minimum achievable number, each resonator must have a bandwidth of $\Delta f_{\scalebox{.6}{\mbox{BW}}}\geq 2 n_{\Phi_0} f_r= 2$\,MHz (equation \ref{eq:shannon}). The current ADCs used for MKIDs and microwave SQUIDs applications have a bandwidth $f_{\scalebox{.6}{\mbox{BW}}}$ of 500\,MHz~\cite{ARCONS}. HOLMES will use a similar board (more details on Section \ref{sec:1000pixels}). By using the relationship \ref{eq:n-pixel} and considering $f_{\scalebox{.6}{\mbox{BW}}}=500$\,MHz, $f_{\scalebox{.6}{\mbox{ramp}}} = 500$\,kHz, $n_{\Phi_0}=2$ and a spacing between resonances around $S=15$ ($S\ge g_f\Delta f_{\scalebox{.6}{\mbox{BW}}}$, equation \ref{eq:space}, with $\Delta f_{\scalebox{.6}{\mbox{BW}}}=2$\,MHz and $g_f=7.5$ for the guard factor), the HOLMES multiplexing factor results 36 pixels per ADC board. This number can be rounded to 32 in order to match the number of resonators per multiplexer chip (more details in Section \ref{sec:mux}). The tipycal RF bandwidth for the HEMT amplifiers selected for HOLMES is in the range from 4 to 8\,GHz~\cite{LNF_LNC4_8C}, this means that a single HEMT can amplify 4000\,MHz/500\,MHz=8 ADC boards, each one up-converted in a different frequency reange, in order to cover the entire HEMT bandwidth. To cover the total 1024 pixels expected for HOLMES, 4 HEMT amplifiers are needed for a total of 32 ADC boards.     

\section{HOLMES multiplexer prototypes}\label{sec:mux}
Multiple generations of 33-channel multiplexers have been fabricated and tested~\cite{MATES_THESIS} at the National Institute for Standard and Technology (NIST, Boulder, Co, USA). Connecting in series few 33-channel chips with different frequency bands allows to increase the multiplexing factor, creating sort of daisy chain~\cite{MATES_SEMINAR}. HOLMES will use 33-channel multiplexer chips designed and provided by NIST and optimized to match the experiment requirements.

The NIST microwave multiplexer consists of 33 quarter-wave coplanar waveguide (CPW) microwave resonators made from a 200\,nm thick Nb film deposited on high-resistivity silicon ($\rho >10\,\mbox{k}\Omega\cdot$cm). A coplanar waveguide consists of a center conductor with ground plane on both sides separated by a gap. The simplicity of this geometry does not require dielectric between the conductors, avoiding a significant mechanism of loss and source of two-level-system noise. Resonators are evenly spaced across the chip. Each one has a trombone-like shape with slightly different length. Resonators couple capacitively to the feedline that runs along the top of the chip, and inductively to their rf-SQUIDs (bottom of the chip). This SQUID loop is a second-order gradiometer consisting of four parallel lobes arranged like a clover-leaf. The NIST process for creating Josephson junctions begins with a trilayer deposition in vacuum~\cite{MATES_THESIS}: deposit niobium (200\,nm), deposit a thin layer of aluminum (7\,nm), flow oxygen to oxidize it, and finalization with more niobium (120\,nm). The entire wafer thus begins as a Josephson junction. The top two layers are etched away over most of the wafer, leaving isolated junction pillars. An additional niobium layer subsequently connects these junctions to SQUID loops

\begin{figure}[!t]
\centering 
\includegraphics[width=0.9\textwidth,clip]{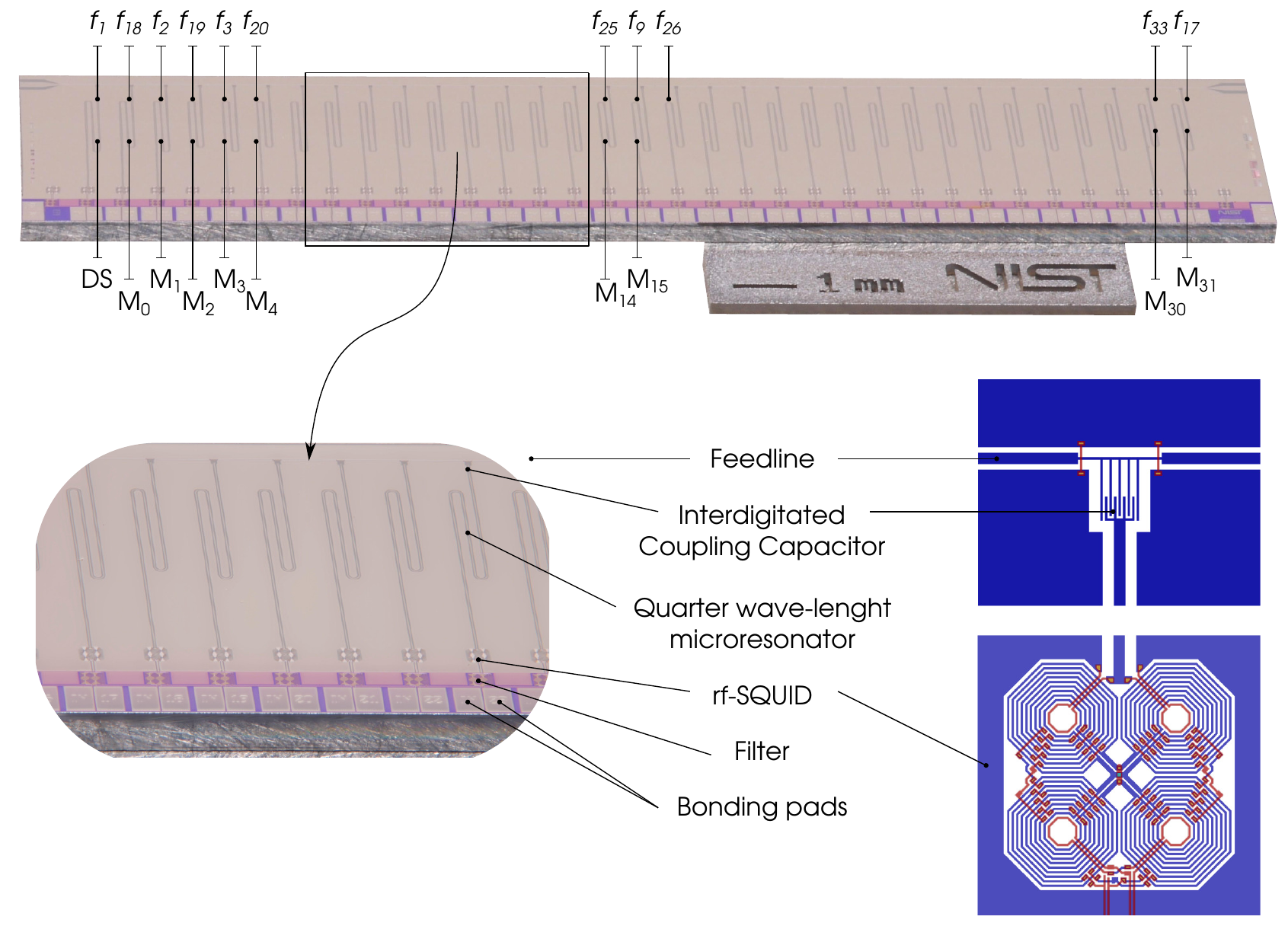}
\caption{\label{fig:umux16a} (top) Multiplexer chip $\mu$mux16a with the array of rf-SQUIDs coupled to the quarter wavelength resonators (trombone-like shaped). (bottom) Enlarged view of the multiplexer highlighting its main parts. Pictures by courtesy of NIST Quantum Sensor Group.}
\end{figure}

The first micro-wave multiplexer specifically designed for HOLMES (code name $\mu$mux16a) is showed in figure \ref{fig:umux16a}. Each chip accommodates 33 quarter-wave microresonators. For convention, the first microresonator on the left is labeled as DS (Dark SQUID), while the other are denoted by an index that monotonically increases, along the direction of the feedline (left-to-right), from 0 to 31 ($M_i$ in figure \ref{fig:umux16a}, top). The resonant frequencies do not follow this order but they are designed to be distributed as follows: $f_1$, $f_{18}$, $f_{2}$, $f_{19}$, ...,  $f_{16}$, $f_{33}$ and $f_{17}$, ($f_j$ in figure \ref{fig:umux16a}, top). In this way two neighbors are well separated in frequency limiting cross talk effects. 

The coupling capacitors are designed for 2\,MHz bandwidth while the rf-SQUID-resonator coupling is designed for 1.5\,MHz peak-to-peak frequency shift. The space between two adjacent-in-frequency resonators is designed around $S=15$\,MHz, that means a guard factor of $g_f=7.5$. The resonance depth is designed to be greater than 10\,dB. The bond pads at the bottom of the chip allow to couple the multiplexer to the detector array. Chips were designed and fabricated at NIST and then characterized at the cryogenic laboratory of Milano-Bicocca. In Milano the multiplexers were cooled to cryogenics temperatures by a pulse tube assisted dilution refrigerator (Oxford Triton 200), instrumented with the necessary superconducting coaxial and twisted lines and electronic components needed to effectively read out the rf-SQUIDs. This cryostat will host the HOLMES experiment in its final configuration. 

\begin{figure}[!t]
\centering\includegraphics[width=1\textwidth,clip]{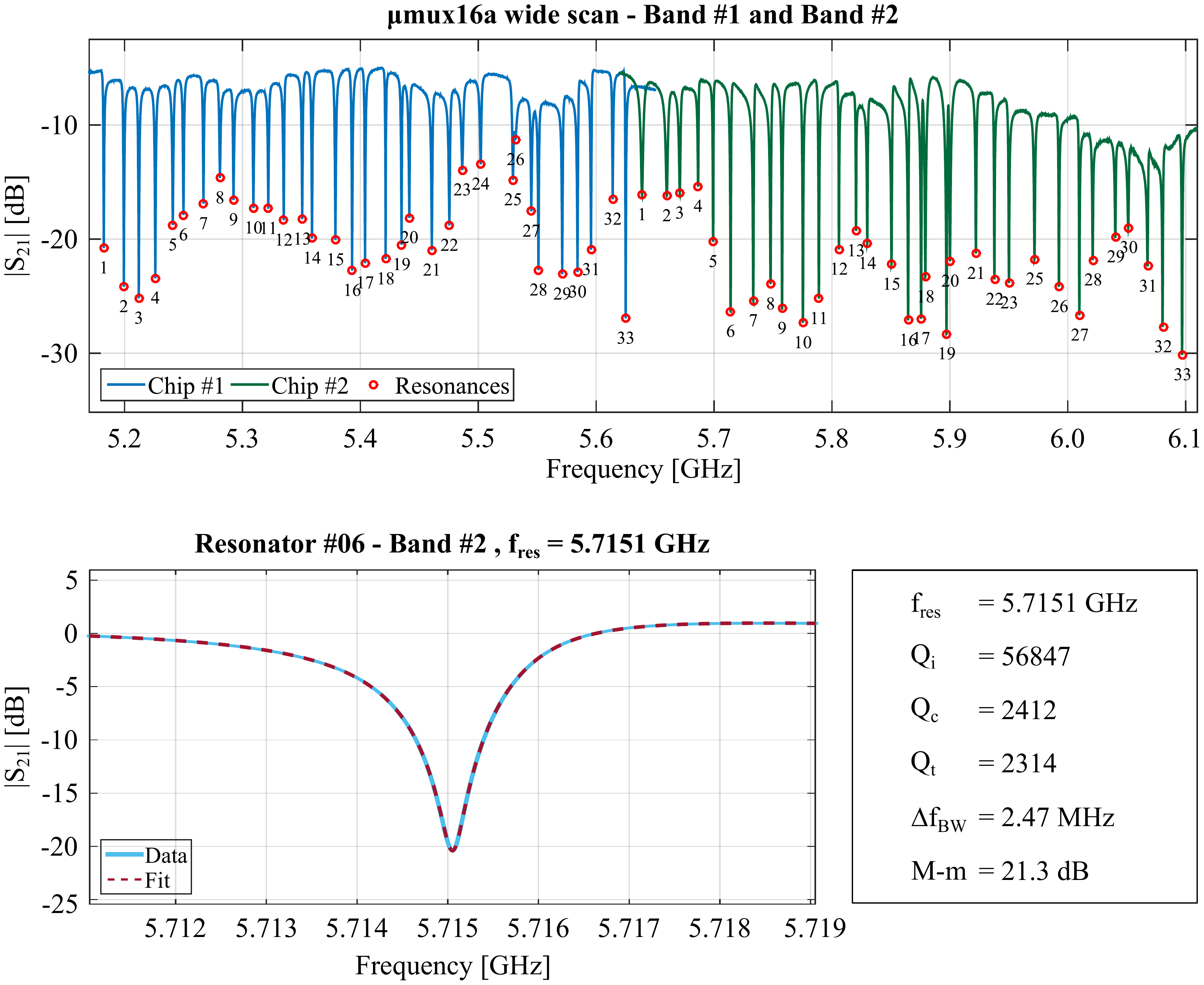}
\caption{\label{fig:umux-scans} (top) Profile of the $S_{21}$ wide scan for two different bands of $\mu$mux16a microwave multiplexer chip. The plot presents a discontinuity and a overlapping around 5.65\,GHz because of the two multiplexer, Band \#1 and  Band \#2, were measured in two different cool downs. (bottom) Profile of a fitted resonance (Resonator \#28, Band \#2) with reported the main extrapolated parameters: $f_{\protect\scalebox{.6}{\mbox{res}}}$, $Q_c$, $Q_i$ , $Q_t$, the resonator bandwidth $\Delta f_{\protect\scalebox{.6}{\mbox{BW}}}$ and the resonator depth computed as maximum ($M$) minus minimum ($m$).}
\end{figure}

The multiplexer chips were characterized by using a remotely controlled vector network analyzer (VNA, model Keysight N9916). The aim was to identify for each resonator the resonance frequency, quality factors, bandwidth and depth. By using the VNA it is possible to perform a scan in frequency of the complex forward transmission parameter $S_{21}$~\cite{Pozar}. Resonances can be identified by analyzing both amplitude and phase of $S_{21}$. Each scan was obtained as average of 30 single scans. This procedure allows to increase the ratio between the measured transmission parameter and the background noise.  

\begin{figure}[!t]
\centering\includegraphics[width=\textwidth,clip]{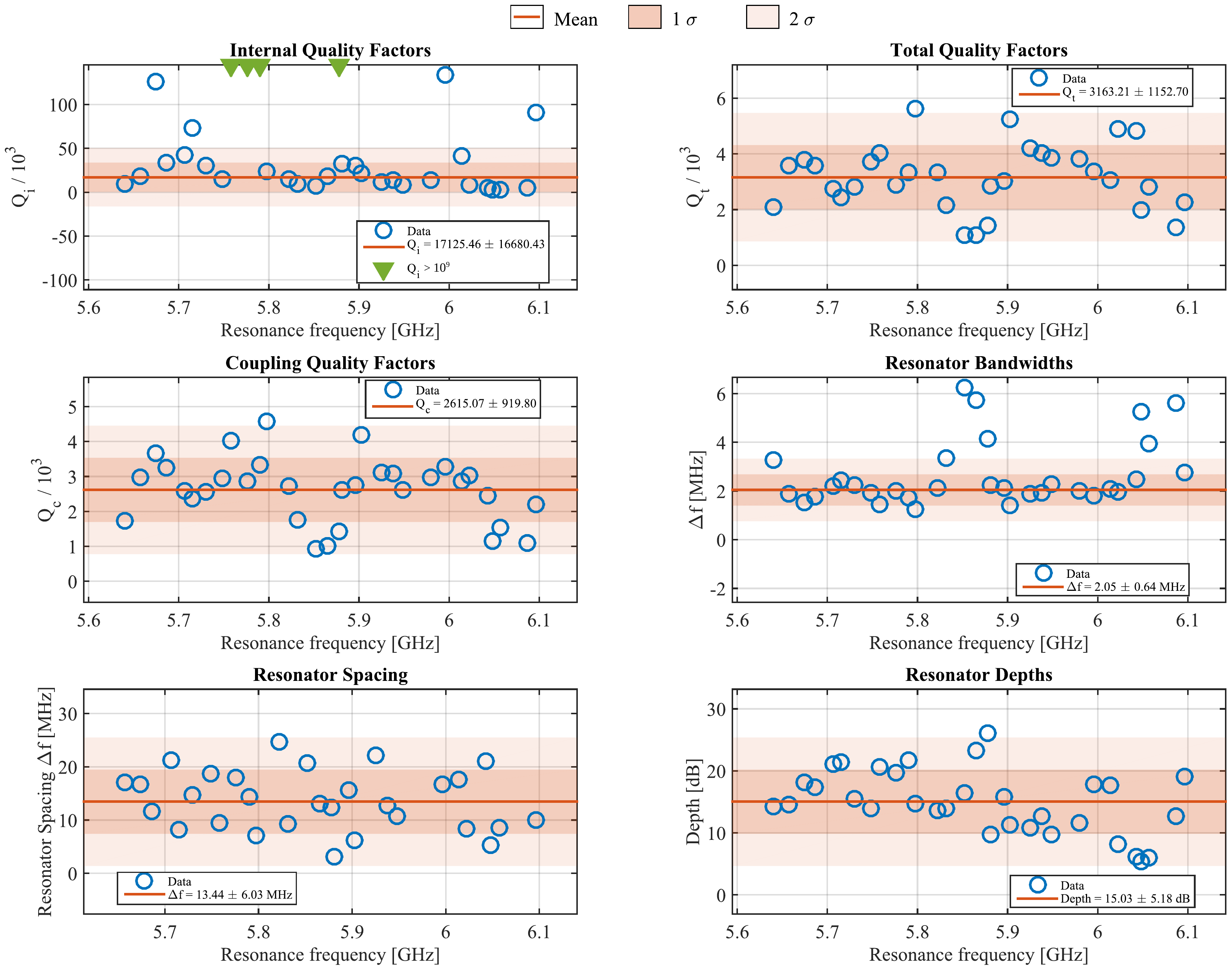}
\caption{\label{fig:fit-report}The multiplexer main characteristic parameters ($Q_i$, $Q_t$, $Q_c$, $\Delta f_{\protect\scalebox{.6}{\mbox{BW}}}$, spacing $\Delta f$ and resonance depth) as a function on the resonant frequency $\Delta f_{\protect\scalebox{.6}{\mbox{res}}}$. The analysis is for the from chip Band \#2, Run \#23, for the other runs similar results were obtained.}
\end{figure}

Two multiplexer $\mu$mux16a prototypes were characterized in Milano and then used to read out the detectors. The first (Band \#1) had frequencies ranging from 5.15 to 5.65\,GHz and the second (Band \#2) from 5.65 to 6.13\,GHz. The first characterization step was a scan across a wide frequency range (around 500\,MHz with resolution of 50\,kHz) necessary to have a rough estimation of the position of the resonances. Afterwards, a fine narrow scan within a range of 10\,MHz (with resolution of 2\,kHz) across each rough resonant frequency was performed for a more precise measurement of the resonance shapes. The fine scans were used to extrapolate the microresonator parameters: the resonant frequency $f_{\protect\scalebox{.6}{\mbox{res}}}$, the quality factors $Q_i$, $Q_t$ and $Q_c$, the resonator bandwidth $\Delta f_{\protect\scalebox{.6}{\mbox{BW}}}$, the spacing between resonators $\Delta f$, and the resonance depth. Each resonance was fitted by using the model illustrated in appendix \ref{sec:appendix-res-model}. The microwave power provided by the VNA during the scans was set to -35,\,dBm that means, considering the fixed attenuators and cryogenic line attenuation, around -80\,dBm at the multiplexer input. This power level was enough to probe the resonators avoiding distortions on the resonator shapes due to non linear effect~\cite{ZmuidzinasMKIDS}. An example of $S_{21}$ wide scan for the two multiplexers (Band \#1 and Band \#2)  is shown in figure \ref{fig:umux-scans} (top).  For the Band \#1 all the 33 resonators are present and visible, while for the Band \#2 the resonator 24 is missing. This issue is probably due to a unpredictable shift of the design value of resonant frequency due to imperfections in the fabrication process that result as a collision of a resonance to the adjacent one. An example of fitted resonance is shown in figure \ref{fig:umux-scans} (bottom), while the main extrapolated parameters as a function of the resonant frequency are reported in figure \ref{fig:fit-report}. The micoresonators are designed in the $Q_c$-limited quality factor condition ($Q_c<<Q_i$). For this reason when the internal quality factor are orders of magnitude greater than the coupling quality factor, the fit procedure returns $Q_t=Q_c$ and $Q_i\to\infty$. This condition is reported in the $Q_i$ vs. $f_{\protect\scalebox{.6}{\mbox{res}}}$ plot (figure \ref{fig:fit-report}, top-left) by denoting this cases with a green filled marker over the maximum plot scale.          

Characterization results showed average values that fulfill the HOLMES requirements: the spacing between adjacent resonances is around $\Delta f=14$\,MHz, the bandwidth is around $\Delta f_{\protect\scalebox{.6}{\mbox{BW}}}=2$\,MHz and resonance depth is around 15\,dB,  greater than 10\,dB. The only issue is the large spread of this values and the fact that this spread changed in every run. Tests and measurements showed that the multiplexer chip response was very sensitive to its electrical connection to the ground-plane. If the resonators chip had not a solid electrical connection to ground this resulted in a degraded resonances shape due to degraded factors of merit. To minimize this effect several and redundant connections have been wired bonded between the chip ground-plane and the copper holder frame and between the ground-plane and the CPW ground. The quality factor is also affected by the number of detectors connected to the multiplexer: by increasing their number, the quality factor gets smaller and smaller, negatively influencing both the bandwidth filling of the chip and the noise contribution of the HEMT amplifier to the demodulated signal. It has been demonstrated by studies carried at NIST that this dependence of the quality factor on the number of connected detectors is due to grounding effects. Nevertheless, it was estimated that these effects can be neglected connecting a number of detectors not greater than four. The NIST group fixed this issue with a new multiplexer chip design ($\mu$mux17a) with optimized ground bondpads position on the chip, and fabricated by employing a new lithography procedure that dramatically improves the resonance frequency placement. The new chip prototypes are currently in use in Milano and the characterization results will be reported in future publications. As reported in Section \ref{sec:noise} another crucial parameter is the read out noise, that must be smaller than the intrinsic noise of the detector. The noise of the multiplexer $\mu$mux16a was measured by using the multiplexing and read out demonstrator here presented and the obtained results are reported in Section \ref{sec:demo}. 

\section{HOLMES detector prototypes}\label{sec:det}
The development of the read out and multiplexing system ran in parallel with the development of the first TES detectors specifically designed for HOLMES. The two-channel system presented in this work, and detailed in the further Sections, was not only used as pure read out demonstrator but also as ready-to-use system for characterize and test these first detector prototypes.   

\begin{figure}[!t]
\centering\includegraphics[width=\textwidth,clip]{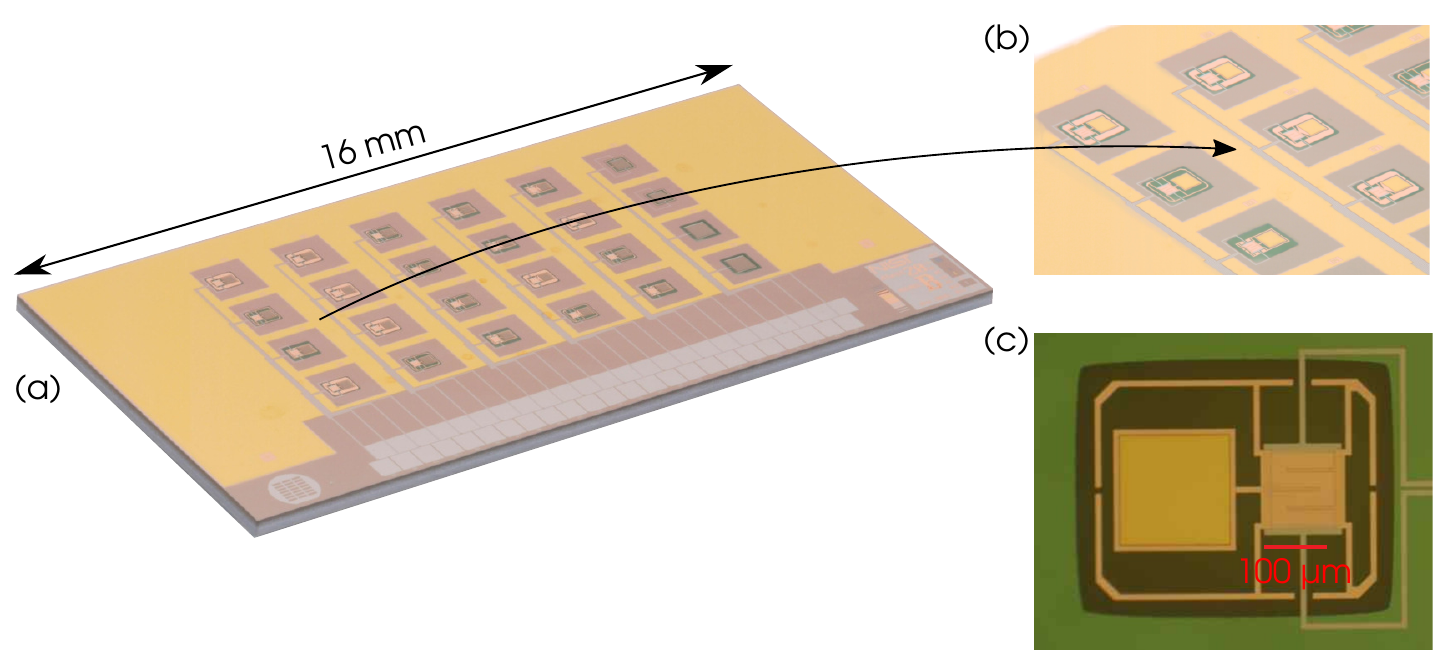}
\caption{\label{fig:dtest20b}(a) Picture of the dtest20b $6 \times 4$ prototype array accommodating different perimeter/absorber configurations, (b) enlarged view of a part of the array, (c) picture from the top of one of the pixel present in the array.}
\end{figure}

The requirement of low pile-up rate sets strict constraints on the detector response. The pulse response must be fast enough to match the rate of the source while a rapid pulse rise can facilitate the pile-up rejection. In a low temperature calorimeter the decay time ($\tau_f$) is set by the ratio between the thermal capacity $C$ of the absorber and the thermal conductance $G$ toward the bath~\cite{McCammon1984,Giachero2017}. The rise time, at the first order, is set by the electrical cutoff of $L_N/R_0$, where $R_0$ is the resistance of the sensor at the
working point, and $L_N$ is a selectable Nyquist inductance connected in series to the TES.

In order to obtain short decay times, the heat capacity must be minimized ($\tau_f\sim C/G$), for example by lowering the absorber size. This conflicts with the minimum absorber thickness required to completely absorb the radiation emitted in the EC decay. Since the fully energy containment is priority requirement, the only way to increase the speed is to increase $G$. A series of different TES prototypes suspended on a SiN membrane were produced and tested at NIST. The selected single pixel design is a $125\times 125\,\mu\mbox{m}^2$ Molybdenum-Copper bilayer TES with two or three normal metal bars built on a SiN membrane. A $200\times 200\times 2\,\mu\mbox{m}^3$ gold absorber is placed alongside the TES to avoid proximity effect between the gold and the sensor itself (side-car design, figure \ref{fig:dtest20b}c). The 2\,$\mu$m-thick gold absorber ensures a 99.99\% (99.73\%) probability of stopping the electrons (photons) coming from the decay of \textsuperscript{163}Ho~\cite{HOLMES_EPJC}. The thermal conductance $G$ is increased by the addition of a thermal radiating perimeter that increases the conductance in this 2-d geometry. Test performed at NIST showed that this thermalizing perimeter increases the thermal conductance $G$ without raising the heat capacity above 0.8\,pJ/K. With this technique any $G$ from 40\,pW/K up to 1\,nW/K is achievable~\cite{Bennett2012,Hays-Wehle2016}. The critical temperature $T_C$ resulted around 100\,mK~\cite{Giachero2016}, as designed, and to maximize the performances the detectors have to work at a base temperature around 40-60,\,mK~\cite{Puiu2017}. Two $6 \times 4$ prototype arrays were fabricated at NIST with slight variations in the pixel perimeter/absorber designs (figure \ref{fig:dtest20b}a, code name dtest20b). The goal was to study the different detector responses selecting the best performing solution for the HOLMES requirements, in term of energy and time resolutions. 

The developed arrays were first measured at NIST with standard techniques~\cite{Lindeman2004}. When operated at a bias point that is 20\% of the normal state resistance $R_n$ (9\,m$\Omega$) they showed a temperature sensitivity $\alpha$ of 60, a current sensitivity $\beta$ of 1.8, and an unexplained noise parameter~\cite{unexplained} $M$ of 1.5. The $C$ and $G$ thermal parameters resulted around within the ranges (0.8-1.0)\,pJ/K and (400-600)\,pW/K, respectively, matching the design values. Moreover, by testing the detectors with a more standard Time Domain Multiplexer~\cite{Doriese2016} a set of detectors showed an energy resolution of 4\,eV, compatible with the energy resolution needed for HOLMES, and a rise and decay time around 10-20\,$\mu$s and 100-150\,$\mu$s, respectively. Monte Carlo simulations based on the TES thermal model~\cite{TES_MODEL}, showed that with these time constants, and with a signal sampling rate of at least 500 kHz, it is possible to obtain a time resolution better than 3\,$\mu$s exploiting discrimination algorithms based on Singular Value Decomposition~\cite{Alpert2015} or Wiener filtering~\cite{Ferri2016}. 

A very important goal for HOLMES was to demonstrate that the detector performances obtained at NIST with a single channel TDM can be reached with the microwave multiplexing without spoiling the detector performances.

\section{The 2-Channel Set-up}\label{sec:2ch}
The development of the HOLMES multiplexing and read out system started in 2015~\cite{Giachero2016} with an analog two-channel system and it is currently in progress exploiting digital electronics (Section \ref{sec:1000pixels}). The 2-channel read out system was based on commercial components. The goal of its development was to demonstrate the techniques and, at the same time, to have an independent system ready-to-use for tests and detector characterizations. Moreover, it represented an useful test bench for all the software and algorithms needed for the demodulation processes, setting a baseline for a future multichannel FPGA-based digital read out.    

\begin{figure}[!t]
\centering\includegraphics[width=\textwidth,clip]{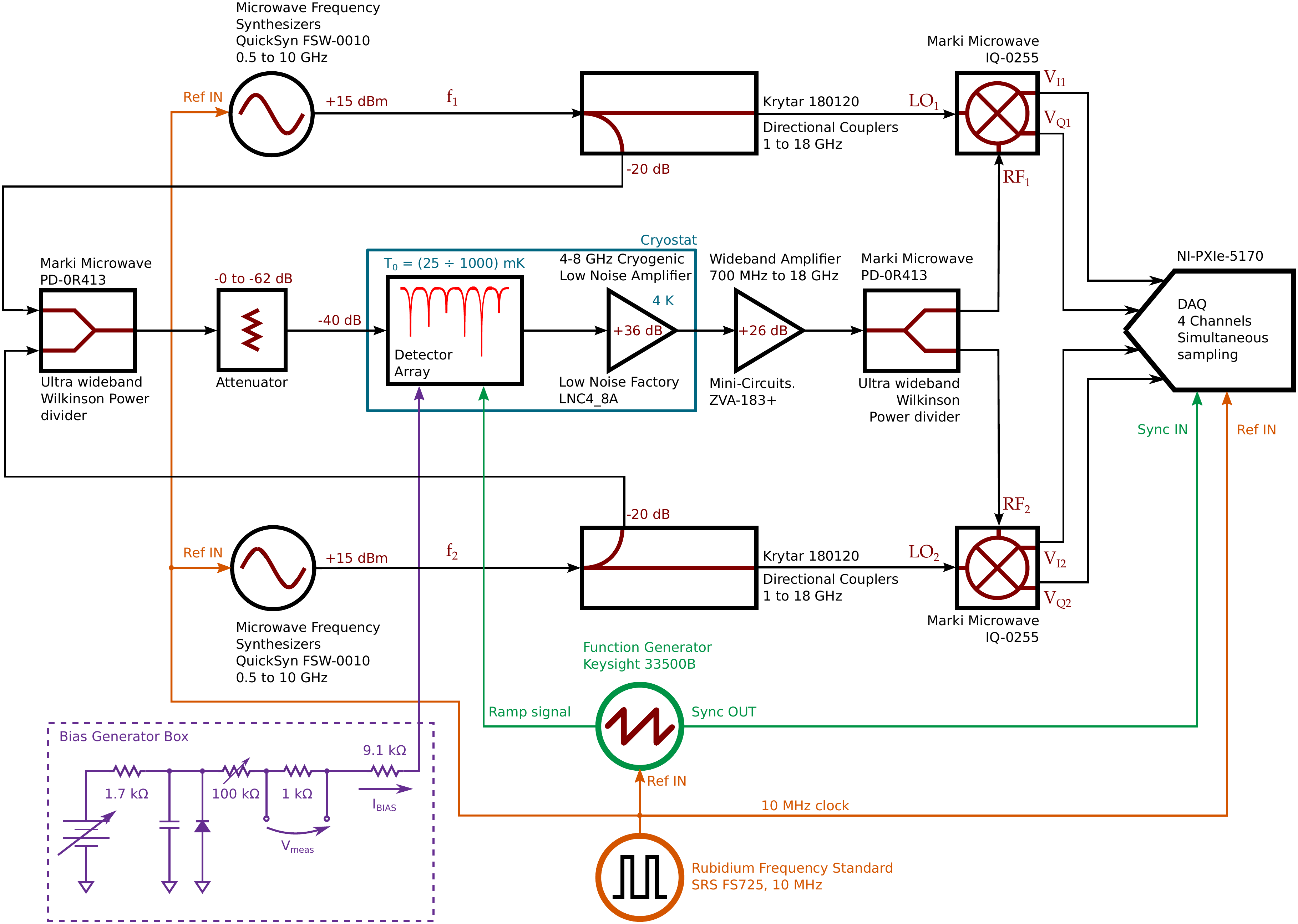}
\caption{\label{fig:homo2ch} Block diagram of the two-channel readout and multiplexing system. Black: RF line, Green: ramp line, Orange: clock line, Violet: bias line.}
\end{figure}

The block diagram of the system is shown in figure~\ref{fig:homo2ch}. Two different probe signals LO$_1$ and LO$_2$, one for each microresonator, were generated directly in RF band by two commercial microwave synthesizers and then split in two. One couple of LO$_1$ and LO$_2$ were merged together by a power coupler, attenuated, and then sent onto the transmission line of the microresonator chip. The resulting forward transmitted waveform was amplified with a cryogenic HEMT amplifier, mounted on the 4\,K stage, and with a room-temperature amplifier. The amplified signal was then split by a power divider in two copies, RF$_1$ and RF$_2$. Homodyne mixing between each copy (RF$_1$ and RF$_2$) and the other copy of the original excitation signals (LO$_1$ and LO$_2$) was accomplished employing two IQ-mixers. The resulting $V_{Q_i}$ and $V_{I_i}$ baseband voltage signals were acquired by a commercial 14 bit-simultaneous-sampling data acquisition board (NI-PXIe-5170) with a sample rate up to $f_{\scalebox{.6}{\mbox{ADC}}}= 250$\,MHz. The total power of the generated signal probe was adjusted by a 0 to 62\,dB step attenuator (with a 1\,dB step) to optimize the signal-to-noise ratio. The final readout powers were around -25/-30\,dBm for each tone that means, considering the fixed attenuators and cryogenic line attenuation, around -70/-75\,dBm for each tone at the multiplexer chip feedline.

The ramp signal was generated by using a remote programmable waveform generator (Keysight 33500B), with a frequency of $f_{\scalebox{0.6}{ramp}}=(400-500)$\,kHz. The amplitude of the ramp is tuned such that it provides $n_{\Phi_0}=2$ of flux per ramp period and modulates $V_{Q_i}$ and $V_{I_i}$ at an oscillation frequency of $f_c=(800-1000)$\,kHz. In order to avoid noise and pick-up due to ground loops the ramp line and the ramp generator were isolated by a coupling transformer~\cite{Transformer}. Introducing a transformer between sawtooth signal generator and the SQUIDs is very effective in suppressing ground loop disturbances and EMI. The waveform generator also generated a Sync OUT TTL signal, synchronized with the the ramp, and acquired by the acquisition board. This signal was used to align the ramp at the beginning of the record and to identify the ramp reset for each ramp period.   

\begin{figure}[!t]
\centering\includegraphics[width=\textwidth,clip]{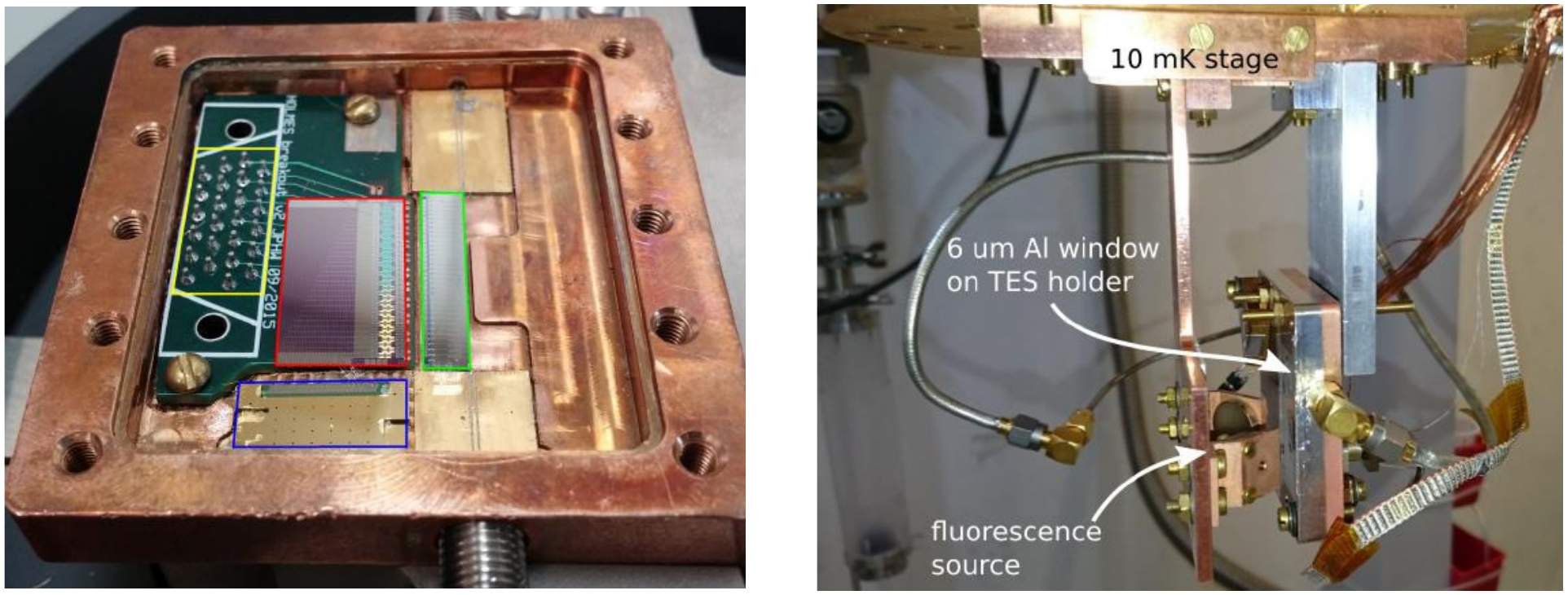}
\caption{\label{fig:holder} (left) The copper holder containing the detectors. In the blue box the TES chip covered with a silicon collimator centered on the sidecar absorbers. In the red box the new interface chip. In green the $\mu$mux16a multiplexer chip. In yellow, a custom PCB board used to provide the ramp line, the bias line and other free input lines. (right) Copper holder facing the fluorescence source Pictures edited from~\cite{PUIU_THESIS}. The copper holder was connectorized with two SMAs for the feedline and a Micro-D for the bias and ramp lines.}
\end{figure}

The TES bias was provided by a custom system composed by a low-noise, low drift, high precision linear voltage reference~\cite{linear} and a voltage divider.  The attenuation was set in order to have a bias current $I_{\scalebox{.6}{\mbox{BIAS}}}$ within the range $(0-2000)\,\mu$A. The output voltage could be tuned by a trimmer that formed a voltage divider. By reading out the voltage across a 1\,k$\Omega$ resistor placed in series at the divider output it was possible to measure the current $I_{\scalebox{.6}{\mbox{BIAS}}}$ provided to the TESs. The linear voltage reference could be powered by a bench power supply or, alternately, with 12\,V batteries. During the I-V curves measurements, where the TESs static response was measured as a function of the bias current, the linear power supply was replaced by a remotely programmable power supply (Agilent 3631A). Finally, to avoid unwanted systematic phase shifts in demodulation due to clock difference all the used instruments (Synthesizers, Ramp Generator and data acquisition) were synchronized by a 10\,MHz Rubidium Frequency Standard (SRS FS725).

The probe signal generated by the homodyne set-up, ran from room temperature to the 4\,K stage through Cu-Be cables, in order to limit the heat load due to the large temperature difference between the stages. After a 20\,dB attenuation at 4\,K in order to match the thermal noise to the temperature difference, the signal was carried to the detectors holder through stainless steel cables which ensures a reduced thermal load on the coldest stage.  Before entering the detector holder, the probe signal is further attenuated by 20\,dB. The copper detector holder accommodated the $\mu$mux16a multiplexer chip, the $6 \times 4$ TES chip and an interface (IF) chip (figure \ref{fig:holder}, left). This last was used to provide a bias shunt resistances of $R_{\scalebox{.6}{\mbox{shunt}}} = 0.33\,\mbox{m}\Omega$ in parallel with each TES and a wirebond-selectable Nyquist inductor $L_N$, in series to set the pulse rise-times ($\tau_r \simeq R_0/L_N$). In the chip the Nyquist inductance options were realized by a N-turn spiral: 0-turn to add no inductance, 6-turn for $\sim$50\,nH of added inductance, 8-turn for $\sim$64\,nH of added inductance, and 10-turn for $\sim$82\,nH of added inductance.


After passing through the multiplexer chip the signal passed through a circulator, configured as isolator to avoid any eventual reflection, and then reached the 4\,K stage through a superconductive Nb coax cable, which ensures lossless transmission. The signal was then amplified at 4\,K by a very low noise HEMT amplifier (Low Noise Factory LNC4\_8A, temperature noise of $T_N=2.5$\,K) and, at room temperature, by a wideband amplifier (Mini-Circuits ZVA-183+). 

\begin{table}[t!]
\centering
\begin{tabular}{|c|ccc|cccc|} 
\hline
\multirow{2}{*}{TES \#}      & \multicolumn{3}{|c|}{Interface (IF) chip} & \multicolumn{4}{|c|}{$\mu$mux16a multiplexer chip} \\
\cline{2-8}
   &     IF &   $N_{\scalebox{.6}{\mbox{turns}}}$ &   $L_N$ [nH] &  $N_{\scalebox{.6}{\mbox{mux}}}$ &  $N_{\scalebox{.6}{\mbox{res}}}$  &  $f_{\scalebox{.6}{\mbox{res}}}$ [GHz]  & $\Delta f_{\scalebox{.6}{\mbox{BW}}}$ [MHz]  \\ 
\hline
  8   &     3  &   6                 &   50     &   4   &   20              &  5.9007                &  1.420    \\
  9   &     4  &   6                 &   50     &   6   &   21              &  5.9223                &  1.799    \\
  11  &     7  &   6                 &   50     &   10  &   23              &  5.9461                &  2.197    \\
  19  &     19 &   6                 &   50     &   20  &   28              &  6.0224                &  1.976    \\ 
\hline
\end{tabular}
\caption{\label{tab:conns} Connections between the TES, IF and multiplexer chips. The reported bandwidth values $\Delta f_{\protect\scalebox{.6}{\mbox{BW}}}$ were obtained with the characterization explained in Section \ref{sec:mux} for the Run \# 26.}
\end{table}

As explained in Section \ref{sec:mux}, to avoid resonances degradation the number of connected detectors to the multiplexer chip was limited to 4. The selected TES were the numbers: 8, 9, 11 and 19 of the dtest20b chip, that presented responses that matched the HOLMES requirements. Since the developed readout system was able to acquire and demodulate only two channels simultaneously, the TESs were acquired as one couple at a time. The connection between the TESs, IF and multiplexer chips are reported in table \ref{tab:conns}.      

A multi line fluorescence source faced to a window on the copper holder was used as calibration source (figure \ref{fig:holder}, right). This was set up using two \textsuperscript{55}Fe emitters faced toward a target of Al, NaCl and CaCO\textsubscript{3} which emitted fluorescence K-$\alpha$ and K-$\beta$ photons. The entrance window of the detector box was covered with a 6\,$\mu$m thick aluminum layer, to prevent the light and the Auger electrons from entering in the box. The TES chip were covered with a silicon micro-machined collimator centered on the sidecar absorbers, in order to avoid signals from radiation hitting the TES or the membrane rather than the absorber.

\begin{figure}[!t]
\centering\includegraphics[width=\textwidth,clip]{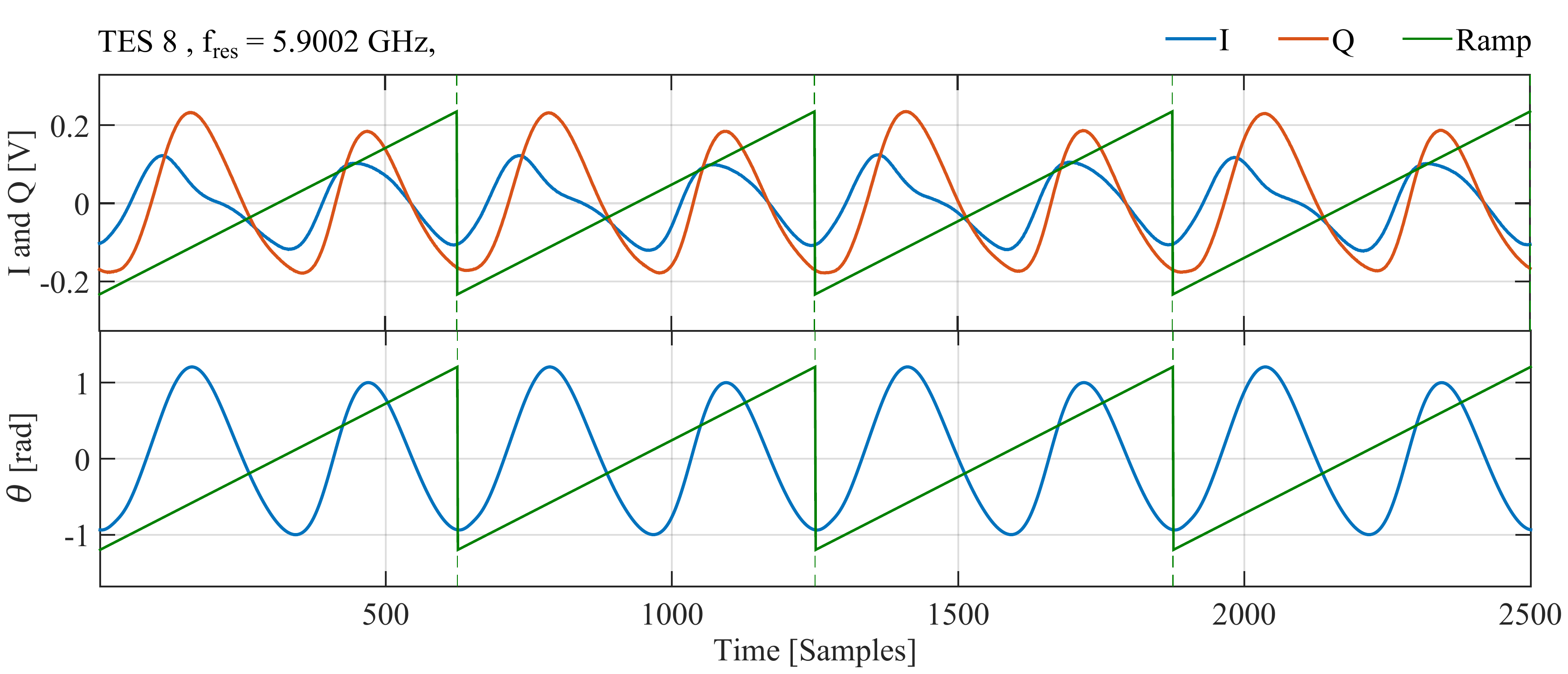}
\caption{\label{fig:free-oscillation} (Top) rf-SQUIDs I and Q free oscillations induced by the ramp modulation. (Bottom) Angle oscillation in the IQ plane obtained after the circle fitting procedure. Data from Run \#27, Measurement \#2, TES \#8}
\bigskip
\centering
\includegraphics[width=\textwidth,clip]{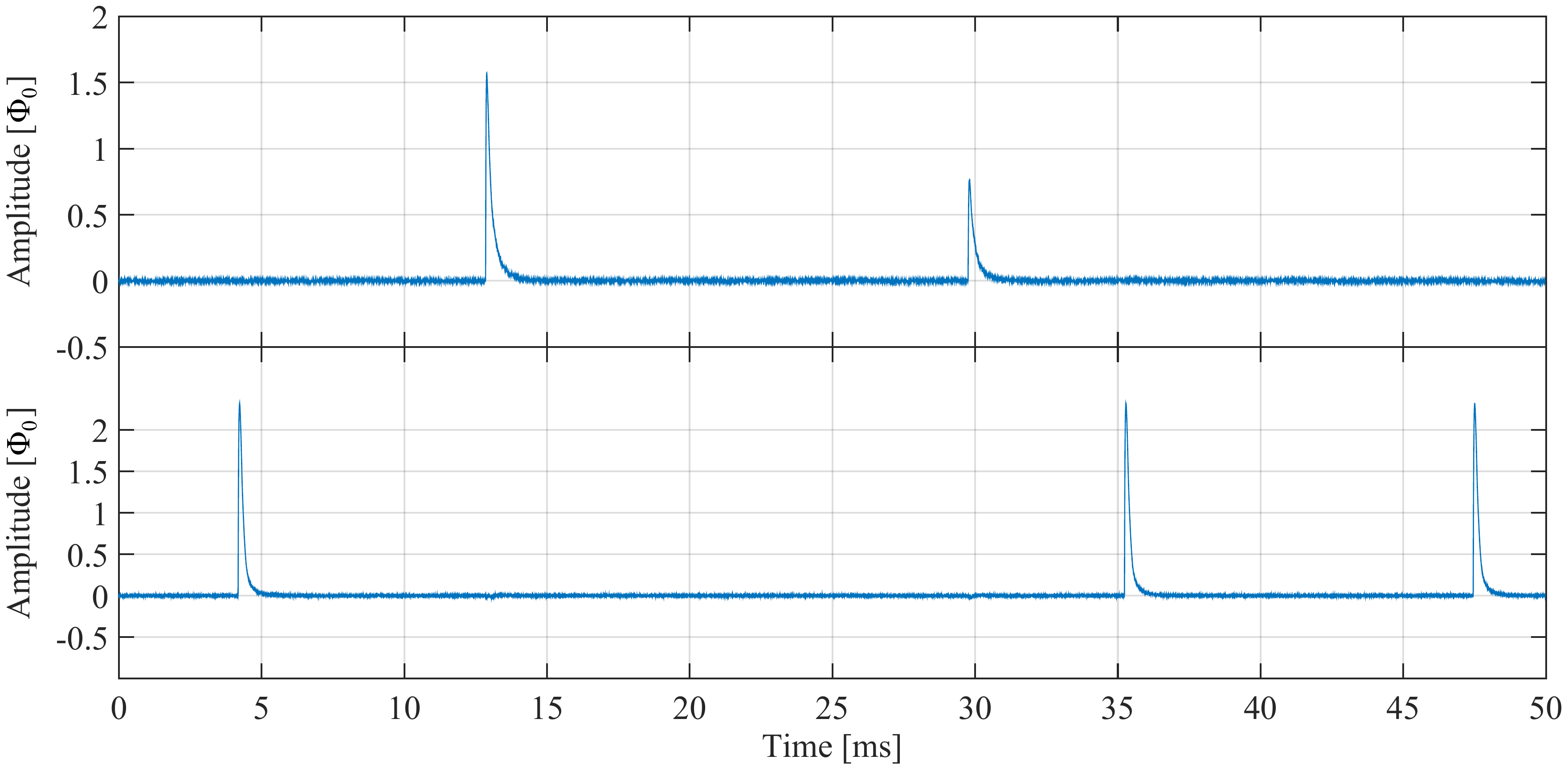}
\caption{\label{fig:2chsim} Simultaneous readout of two Transition-Edge microcalorimeters using microwave rf-SQUID multiplexing. }
\end{figure}

\section{Detector readout demonstration}\label{sec:demo}
The system presented in Section \ref{sec:2ch} was totally configurable by low level software controlled by a high level GUI, both developed in the MATLAB computing environment. The selection of the two TESs to acquire was performed by setting the LO$_1$ and LO$_2$ frequencies: around $f_{\scalebox{.6}{\mbox{LO}}_1}=5.9007$\,GHz and $f_{\scalebox{.6}{\mbox{LO}}_2}=5.9223$\,GHz for TES \#8 and \#9, and around $f_{\scalebox{.6}{\mbox{LO}}_1}=5.9461$\,GHz and $f_{\scalebox{.6}{\mbox{LO}}_2}=6.0224$\,GHz for TES \#11 and \#19. The ramp amplitude ($V_{\scalebox{.6}{\mbox{ramp}}}$) and frequency ($f_{\scalebox{.6}{\mbox{ramp}}}$) were tuned in order to have 2 or 3 flux oscillations per ramp period, while the ADC sample rate was selected to have a number of $V_{Q_i}$ and $V_{I_i}$ samples per ramp adequate to be sensitive to the SQUID oscillations. Since the optical link between the data acquisition crate and the control computer was not optimized to read a high speed simultaneous 4-channel board, each acquisition were performed downloading the entire on-board buffer, with an unavoidable dead-time in between. Working at the maximum ADC sampling rate $f_{\scalebox{.6}{\mbox{ADC}}}=250$\,MHz, with a ramp frequency of  $f_{\scalebox{.6}{\mbox{ADC}}}=500$\,kHz, and acquiring buffer length of around $4.096\times 10^6$ samples, the number of acquired consecutive ramp for each buffer were 8192, that means a record of 8192 demodulated samples, large enough to evaluate the temporal response of the TESs under study.

\begin{figure}[!t]
\centering
\includegraphics[width=\textwidth,clip]{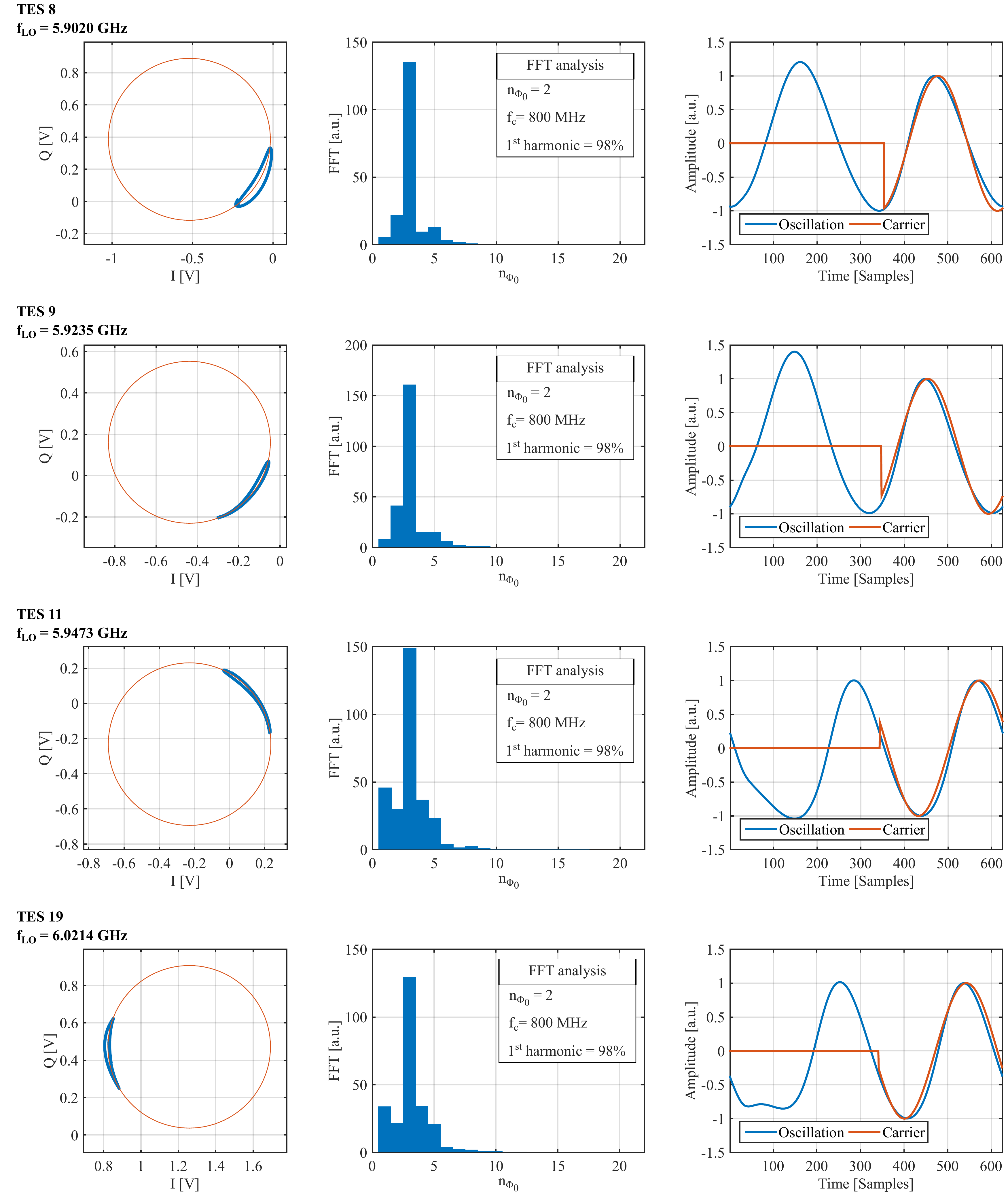}
\caption{\label{fig:free-oscillation-4TES} Example of evaluation of SQUID free oscillation for the four TESs connected to the multiplexing chip. (Left) SQUID Oscillations in the IQ plane with circle fitting. (Center) FFT analysis for the estimation of the SQUID oscillation frequency reported in term of number of flux quanta $\Phi_0$ per ramp period. (Right) Comparison between the SQUID angle oscillation and the sinusoidal carrier used for the ramp demodulation. Data from Run \#27: Measurement \#2 for TES \#8 and \#19 and Measurement \#11 for TES \#9 and \#11.}
\end{figure}

Before the effective acquisition and demodulation processes, the first step was to evaluate the oscillations of the rf-SQUIDs in the absence of TES bias (free oscillations, \ref{fig:free-oscillation}). This step was fundamental to generate the proper sinusoidal signal used as demodulating carrier in the demodulation procedure (as explained in Section \ref{sec:umux}). A number of $V_{Q_i}$ and $V_{I_i}$ oscillations were acquired, divided per ramp period and then averaged obtaining an oscillation arc on the IQ circle (figure \ref{fig:free-oscillation-4TES}, left column), similar to the one shown in figure \ref{fig:IQcircle}. From the arc it was possible to extrapolate the circle center ($I_{c_i}$, $Q_{c_i}$) and the radius ($R_i$) by using a fit procedure based on the K\r{a}sa~\cite{Kasa} or Taubin~\cite{Taubin} algorithms. The oscillating signal $\theta_c(t)$ (figure \ref{fig:free-oscillation}, bottom) is then obtained as arc tangent of the centered $V_{Q_i}(t)$ and $V_{I_i}(t)$ (following equation \ref{eq:demod-theta}, with $\phi_i=0$). Finally, $\theta_c(t)$ is analyzed by using a standard Fourier analysis, identifying the fundamental frequency $f_c$ and, eventually, the first and second order harmonics (figure \ref{fig:free-oscillation-4TES}, center column). The number of flux quanta per ramp $n_{\Phi_0}$ was computed comparing the fundamental and ramp frequencies. The values of $f_{\scalebox{.6}{\mbox{LO}}_i}$ and $V_{\scalebox{.6}{\mbox{ramp}}}$ were set in order to obtain the best fit results. The final obtained values of carrier frequency $f_c$, IQ center ($I_{c_i}$, $Q_{c_i}$) and radius $R_i$ were used to demodulate the signal in the normal data taking (figure \ref{fig:free-oscillation-4TES}, right column). The circle radius, the circle center coordinates and peak-to-peak oscillation for the 4 connected TESs are listed in \ref{tab:carriers}. As expected, the obtained values resulted very similar for the all detectors.     
\begin{table}[!t]
\centering
\begin{tabular}{|c|ccc|cccc|} 
\hline
 TES \#  &  $N_{\scalebox{.6}{\mbox{mux}}}$ &  $N_{\scalebox{.6}{\mbox{res}}}$  &  $f_{\scalebox{.6}{\mbox{LO}}}$ [GHz]  & $I_c$ [V] & $Q_c$ [V] & $R$ [V] & $\theta_{pp}$ [rad] \\ 
\hline
  8   &  4   &   20              &  5.9020                & -0.52  &  0.39  & 0.50 &  0.85 \\
  9   &  6   &   21              &  5.9235                & -0.44  &  0.16  & 0.39 &  0.97 \\
  11  &  10  &   23              &  5.9461                & -0.23  & -0.23  & 0.46 &  1.00 \\
  19  &  20  &   28              &  6.0224                &  1.26  &  0.47  & 0.44 &  0.89 \\
\hline
\end{tabular}
\caption{\label{tab:carriers} Fit results obtained after the circles characterization procedure.}
\end{table}

For each ramp period present in each ADC acquisition record, the $V_{Q_i}(t,\phi)$ and $V_{I_i}(t,\phi)$ signals were centered by using the center coordinates ($I_{c_i}$, $Q_{c_i}$), then the oscillating response $\theta(t,\phi=0))$ were extrapolated following the equation \ref{eq:demod-theta}, and last the phase angle $\phi$ is extrapolated following the equation \ref{eq:demod-phi} applying a sine and cosine demodulation carriers with frequency $f_c$. By biasing the TESs  and then by performing the demodulation it was possible to observe events due to the calibration source in real time in both the acquired channel (figure \ref{fig:2chsim}). The system was able to work in three different modes: IV curve tracer, power noise meter, and X-ray spectrometer.

\subsection{IV curve measurements}    
The I-V curve measurement of a TES detector is a fundamental procedure that allows to extract its main parameters (operating current $I_{\scalebox{.6}{\mbox{TES}}}$, voltage $V_{\scalebox{.6}{\mbox{TES}}}$, and resistance $R_{\scalebox{.6}{\mbox{TES}}}$, or $R_{0}$) as a function of the applied bias current  ($I_{\scalebox{.6}{\mbox{BIAS}}}$). The I-V curve measurements also establishes a relation between the voltage across the TES ($V_{\scalebox{.6}{\mbox{TES}}}$) and the current through the TES ($I_{\scalebox{.6}{\mbox{TES}}}$). This information is particularly important in the transition region where the TES is normally biased. The analysis of IV curves taken at multiple bath temperatures can also provides information about the thermal conductance of the TES to the thermal bath~\cite{TES_MODEL}. 

\begin{figure}[!p]
\centering\includegraphics[width=\textwidth,clip]{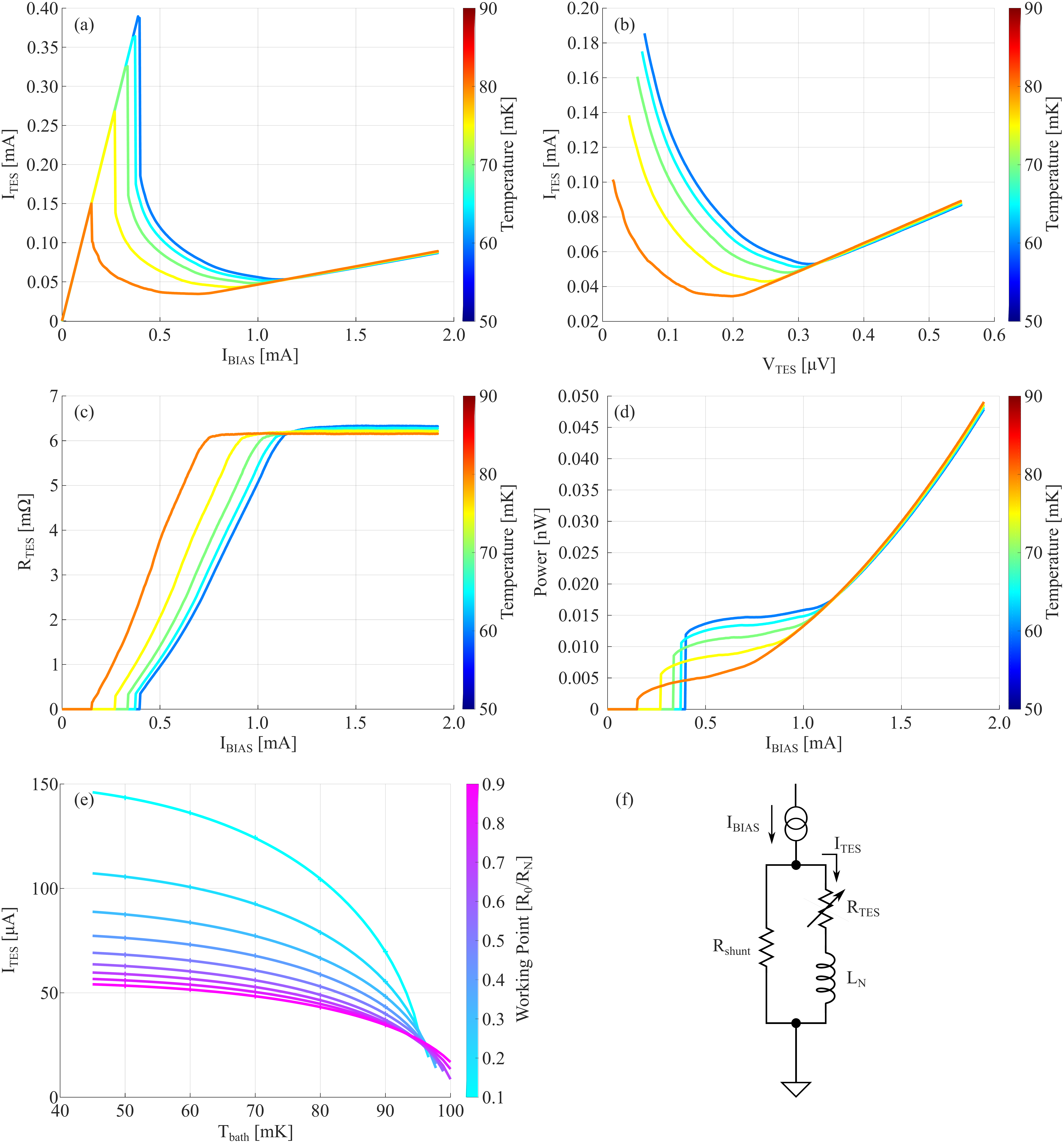}
\caption{\label{fig:ivcurves} IV curve measurement for the TES \#9, Run \#26. The curves (a), (b), (c) and (d) were taken at various bath temperatures between 40\,mK and 80\,mK. From this curves is possible to extrapolate the SQUID transduction gain $M$ and the TES normal state resistance $R_N$. The curves (e) were computed starting from the previous plot for different working point in the range between 20\% and 80\%, and then used to extrapolate the TES thermal parameters $G$, $n$ and $T_c$. The schematic (f) represents the simplified bias circuit.}
\end{figure}

The I-V curves measurements were performed with the bias circuit feeded by a programmable power supply, remotely controlled by the read out control system, instead of the linear voltage reference. To avoid the introduction of possible EMI disturbances and ground loops during the measurements, the calibration curve $V_{\scalebox{.6}{\mbox{supply}}}\leftrightarrow I_{\scalebox{.6}{\mbox{BIAS}}}$ were previously measured by reading out the voltage across a 1\,k$\Omega$ resistor, as discussed in Section \ref{sec:2ch}. In figure \ref{fig:ivcurves}(f) the simplified bias circuiut is shown. 

A typical TES I-V curve is divided into three regions: the normal region, the transition region, and the superconducting region. To obtain a TES I-V curve at a bath temperature $T_0$, lower than the TES critical temperature, the $I_{\scalebox{.6}{\mbox{BIAS}}}$ current ranged from an initial value, high enough so that the TES was in its normal state, down to zero, scanning all the three regions.  The set up was able to measure the IV curves for two TESs simultaneously. For each $I_{\scalebox{.6}{\mbox{BIAS}}}$ the related input fluxes $\Phi$ were obtained as the average of a demodulated record 1024 samples long. Since it was not possible to turn the calibration source off, a rejection algorithm was applied to discard baselines with events due X-ray interactions or with a shape non compatible with a flat stable condition. The system measured the ($\Phi$ vs. $I_{\scalebox{.6}{\mbox{BIAS}}}$) curve. From this curve it was possible to extrapolate the SQUID transduction gain $M=\Phi/I_{\scalebox{.6}{\mbox{TES}}}$ as linear fit in the superconducting section where $R_0=0$ and $I_{\scalebox{.6}{\mbox{TES}}}=I_{\scalebox{.6}{\mbox{BIAS}}}$. 
This found value was used to transform the ($\Phi$ vs. $I_{\scalebox{.6}{\mbox{BIAS}}}$) curve in  ($I_{\scalebox{.6}{\mbox{TES}}}$ vs. $I_{\scalebox{.6}{\mbox{BIAS}}}$). Then the ($V_{\scalebox{.6}{\mbox{TES}}}$ vs. $I_{\scalebox{.6}{\mbox{TES}}}$) and ($R_0$ vs $I_{\scalebox{.6}{\mbox{BIAS}}}$) curves were obtained by using the Thevenin equivalent circuit~\cite{TES_MODEL}. Finally the (Power vs  $I_{\scalebox{.6}{\mbox{BIAS}}}$) curve can be obtained as $P_J=V_{\scalebox{.6}{\mbox{TES}}}I_{\scalebox{.6}{\mbox{TES}}}$. In figure \ref{fig:ivcurves} an example of an IV curve measurement for the TES \#9 is reported. The other detectors showed similar behaviours. 

In a TES detector, the heat loss through the weak thermal link toward the bath can be expressed by the power law $P_{\scalebox{.6}{\mbox{BATH}}}=K(T_{\scalebox{.6}{\mbox{BATH}}}^n-T^n)$~\cite{TES_MODEL}, where $n$ is the thermal conductance exponent, which depends on the dominant thermal transport mechanism, $K=G/(nT^{n-1})$ is a constant that depends on the geometry and material properties of the supporting structure ($G$ is the thermal conductance between the TES and the thermal bath), and $T$ is the temperature at which the TES is kept by the bias current (typically around the TES critical temperature). Since the Joule heating term is equal to the power that flows through the thermal link, it is possible to write

\begin{equation}\label{eq:G}
V_{\scalebox{.6}{\mbox{TES}}}\,I_{\scalebox{.6}{\mbox{TES}}}=R_0\,I_{\scalebox{.6}{\mbox{TES}}}^2=\cfrac{G}{nT^{n-1}}\,\left(T_{\scalebox{.6}{\mbox{BATH}}}^n-T^n\right)~\quad\Rightarrow\quad~
I_{\scalebox{.6}{\mbox{TES}}}=\sqrt{
\cfrac{1}{R_0}\,\cfrac{G}{nT^{n-1}}\,\left(T_{\scalebox{.6}{\mbox{BATH}}}^n-T^n\right)
}
\end{equation}

Considering a given bias working point in TES resistance (expressed as percentage of the normal resistance $R_N$) it is possible to plot the related $I_{\scalebox{.6}{\mbox{TES}}}$ against bath temperature $T_{\scalebox{.6}{\mbox{BATH}}}$. By fitting the obtained behaviour with the relationship \ref{eq:G} it is possible to estimate the values of $T$, $n$ and $G$. In figure \ref{fig:ivcurves}(e) and table \ref{tab:Gfit} the obtained fit results for the four measured TESs are reported. The selected working point was $R_0/R_N=80\%$.  All the TESs showed homogeneous fit values and the obtained results for $G$ and $T=T_c$ resulted quite similar to the value measured at NIST: $G=570$\,pW/K and $T_c=100$. Also the thermal conductance exponent $n$ resulted compatible with the theoretical one, since the thermal model indicates a value within the range from 3 to 4~\cite{TES_MODEL}. The four SQUIDs showed also compatible transduction gains values around 86\,$\Phi_0$/mA. The characterization performed at NIST was based on a TDM system and, in general, with a set up configuration totally different from the one used in Milano. The similarity of the results is a hint of the goodness of the developed microwave multiplexing system, ensuring also the absence of performance degrading problems due to the read out.

\begin{table}[!t]
\centering
\begin{tabular}{|c|ccc|cc|} 
\hline
 TES \#  &  $G$ [pW/T]  &  $n$  & $T$  [K] & $M$ [$\Phi_0$/mA]& $R_{N}$ [m$\Omega$] \\ 
\hline
8   & 591$\pm$10   & 3.96$\pm$0.05   & 99.4$\pm$0.1   & 85.99$\pm$0.14  & 6.51$\pm$0.10\\ 
9   & 626$\pm$16   & 3.79$\pm$0.12   & 101.6$\pm$0.3  & 86.00$\pm$0.22  & 6.26$\pm$0.07\\ 
11  & 523$\pm$18   & 3.50$\pm$0.14   & 100.4$\pm$0.4  & 86.11$\pm$0.18  & 6.44$\pm$0.07\\ 
19  & 568$\pm$13   & 3.81$\pm$0.12   & 98.6$\pm$0.3   & 86.07$\pm$0.21  & 6.50$\pm$0.09\\ 
\hline
\end{tabular}
\caption{\label{tab:Gfit} Fit results obtained from the IV curve analisys for the TES \#8, \#9, \#11 and \#19 from Run \# 26.}
\end{table}

\subsection{Noise measurements}

\begin{figure}[!t]
\centering\includegraphics[width=\textwidth,clip]{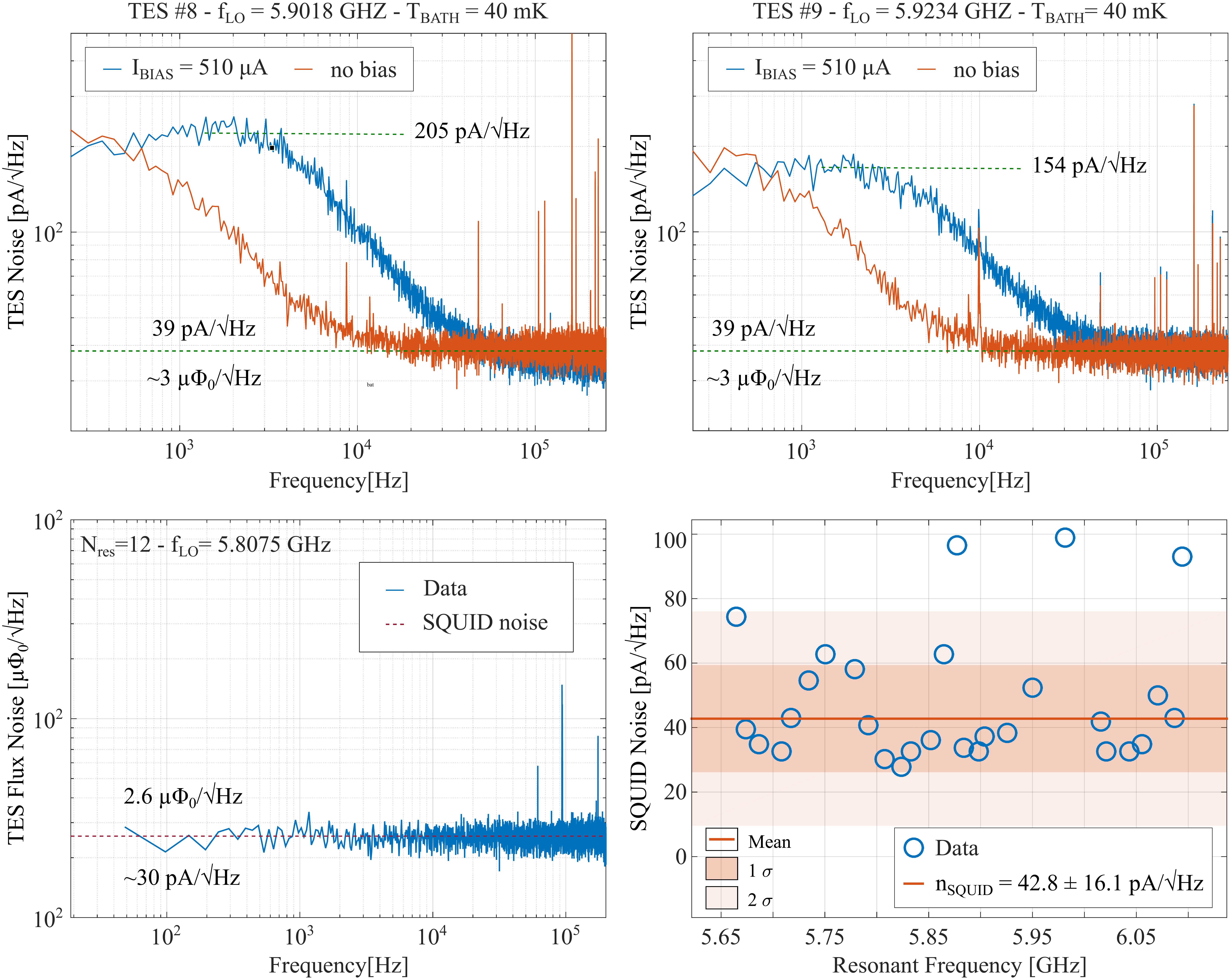}
\caption{\label{fig:noise} (top) Example of amplitude noise spectral density computed online for the TES \#8 and \#9 in case of no bias and bias set to $R_0/R_N=20\%$. (bottom-left) Example of amplitude noise spectral density computed online for the resonator $N_{\protect\scalebox{.6}{\mbox{res}}}=12$. (bottom-right) Amplitude noise spectral density measured for 29/33 rf-SQUIDs present on the multiplexer chip $\mu$mux16a. The transduction gain used to transform the input flux noise in current noise is around 86\,$\Phi_0$/mA, as measured in table \ref{tab:Gfit} for the rf-SQUIDs with connected TES}
\end{figure}

The developed 2-channel system was used to compute the detector noise spectral density in real-time. By using the live computation it was possible to optimize both the hardware part (connections and groundings) as well as the demodulation parameters before starting the measurement phase.  The spectral density was computed by averaging a number of spectra obtained by computing a FFT on demodulated records. As for the IV curve measurements, when the TESs were biased, a rejection algorithm was applied to discard record with pulses or with shape non compatible with a stable condition. The number of averages were configurable (typical value $\sim$ 50) and the number of points per records could be set from 1024 (spectral resolution of $f_{\scalebox{.6}{\mbox{min}}}=488$\,Hz at $f_{\scalebox{.6}{\mbox{ramp}}}=500$\,kHz) to 16384 (spectral resolution of $f_{\scalebox{.6}{\mbox{min}}}=30$\,Hz at $f_{\scalebox{.6}{\mbox{ramp}}}=500$\,kHz). The amplitude spectral density (ASD) was computed in $\Phi_0/\sqrt{\mbox{Hz}}$. To converted the spectra in pA/$\sqrt{\mbox{Hz}}$ the $M$ factors obtained in the IV curves measurements, and reported in table \ref{tab:Gfit}, were used. The noise measurements were performed also for all the rf-SQUIDs without any TES connected to the input coil. This type of measurements gave information on the noise due to the read out only.   

In figure \ref{fig:noise} the noise spectra for TESs \#8 and \#9 and, for the sake of example, the spectrum for an open SQUID are shown. TES \#11 and \#19 presented similar spectra. The TESs showed a noise spectra density of $200\,\mbox{pA}/\sqrt{\mbox{Hz}}$ at most, as expected. With this noise level the best achievable resolution is 4\,eV at 6\,keV. In figure \ref{fig:noise} (bottom-right) the rf-SQUID noise levels measured for 29 over 33 rf-SQUIDs on the $\mu$mux16a chip are showed. The SQUIDs for which no data are reported were not measurable due to difficulties in demodulating the signals from the correspondent resonators. The noise spectral density of the SQUIDs varied between 2\,$\mu\Phi_0/\sqrt{\mbox{Hz}}$ ($\sim23\,\mbox{pA}/\sqrt{\mbox{Hz}}$) and 8\,$\mu\Phi_0/\sqrt{\mbox{Hz}}$ ($\sim93\,\mbox{pA}/\sqrt{\mbox{Hz}}$). In order to not become a significant source in energy resolution degradation, the ideal SQUID noise should not be above 3\,$\mu\Phi_0/\sqrt{\mbox{Hz}}$. A possible reason for the measured spread could be the fluctuation of the coupling to the feedline described in Section \ref{sec:mux}. In fact, the degradation of the resonance factors of merit made the HEMT noise dominant increasing the total read out noise. In the multiplexer chip currently in use ($\mu$mux17a) this issue was fixed achieving a read out noise spectral density level of 2\,$\mu\Phi_0/\sqrt{\mbox{Hz}}$ with a very low spread around the 10\%~\cite{128Mux}. Results of the noise measured with this new chip with HOLMES detectors coupled is reported in a recent publications~\cite{HOLMES_EPJC}.   
  
\subsection{Energy calibration measurements}
In order to verify the microwave multiplexing readout performances in terms of energy resolution with respect to the TDM read out a calibration measurement was performed. The amplitude evaluation of the pulse events was done by maximizing the signal-to-noise ratio by means of optimum filtering techniques~\cite{Gatti_Manfredi}. 

For the two selected TESs the optimal working point was chosen and  they were illuminated with the fluorescence source previously described. Each measurement consisted of a data-taking few days long. TESs data were continuously streamed into the control PC where the on-line ramp demodulation procedure was performed, obtaining a demodulated record sampled at $f_{\scalebox{.6}{\mbox{ramp}}} = 500$\,kHz and 8192 points long. For each channel, data were saved to disk only when a 2nd-order derivative trigger fired. The system recorded 1024 samples ($\sim$2\,ms) for each event, where the first 128 or 256 samples corresponded to a pre-triggered timing region. To derive a noise power spectrum, needed for the optimum filtering, a large set of randomly-acquired baselines (waveforms of pure noise) were collected from demodulated records where no pulse events were triggered. The optimum filter and the average pulse were computed offline by the analysis software.

\begin{figure}[!t]
\centering\includegraphics[width=\textwidth,clip]{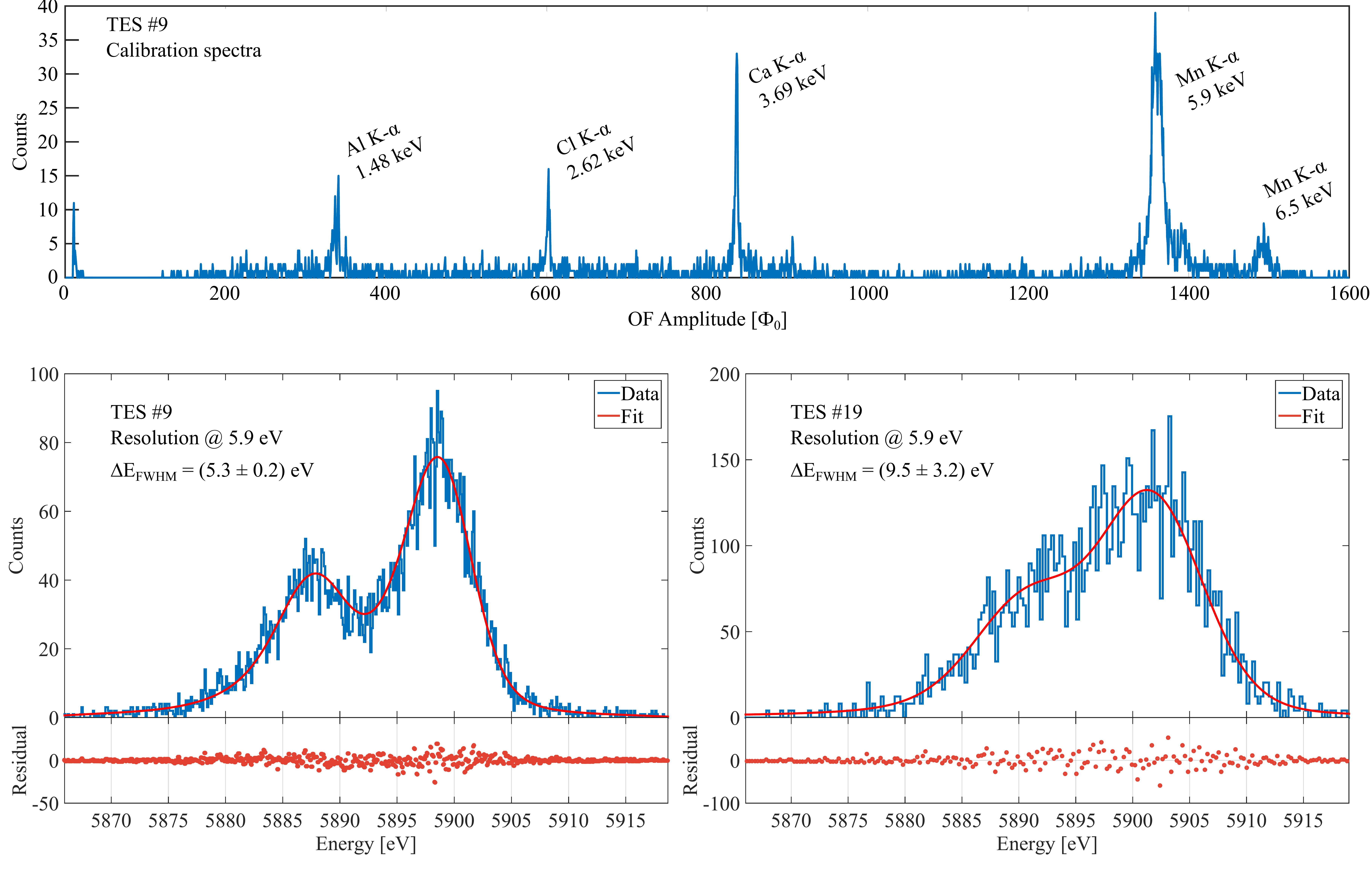}
\caption{\label{fig:res} (top) Calibration spectrum with the composite X-ray fluorescence source obtained from raw data without the application of signal quality and shape cuts. The spectral lines of Al, Cl, Ca and Mn can be recognized. Edited from \cite{PUIU_THESIS}. (botton-left) Obtained energy resolution for the TES \#9 at the \textsuperscript{55}Mn K$\alpha$ peaks ($R_0/R_N=16\%$ and $T_0=40$\,mK)}. 
\end{figure}

An example of calibrated energy spectrum arising from a composite X-ray fluorescence source evaluated with the optimum filter for the TES \# 9 is shown in figure \ref{fig:res} (top). The other detectors showed similar calibration spectra. The detector resolutions were evaluated from the known energy peaks of the spectrum. Since the events from the \textsuperscript{55}Mn K$\alpha$ resulted with higher statistic, the detector resolutions were estimated on this peak. The \textsuperscript{55}Mn K$\alpha_1$/K$\alpha_2$ peak (5.89875\,keV and 5.88765\,keV) were fitted with the analysis procedure previously developed for the MIBETA experiment~\cite{peakfit}. In figure \ref{fig:res} (bottom) two examples of fit result are reported. The best performing detector was the TES \#9 with a FWMH resolution  of $(5.3 \pm 0.2)$\,eV at 5.9\,keV. The best energy resolutions for the four measured TESs are reported in table \ref{tab:ress}. These results represented a good achievement considering that they were obtained with a brand new read out system based on a cutting edge multiplexing technique. Nevertheless, considering that the computed signal-to-noise ratio were around or above 1000 at 6\,keV, these values exceeded what expected by the optimal filter analysis (table \ref{tab:ress}). Furthermore the TES \#9 showed an experimental energy resolution of about 4\,eV when acquired with a TDM system at NIST. This mismatch was neither ascribed to the detector or to the read out system, but to thermal instability on the long time scale that causes drifts of the detector response. To solve this issue the decoupling between the Mixing Chamber stage of the cryostat and the detector box have been improved and temperature control, performed by an active resistance bridge, have been implemented.

\begin{figure}[!t]
\centering\includegraphics[width=\textwidth,clip]{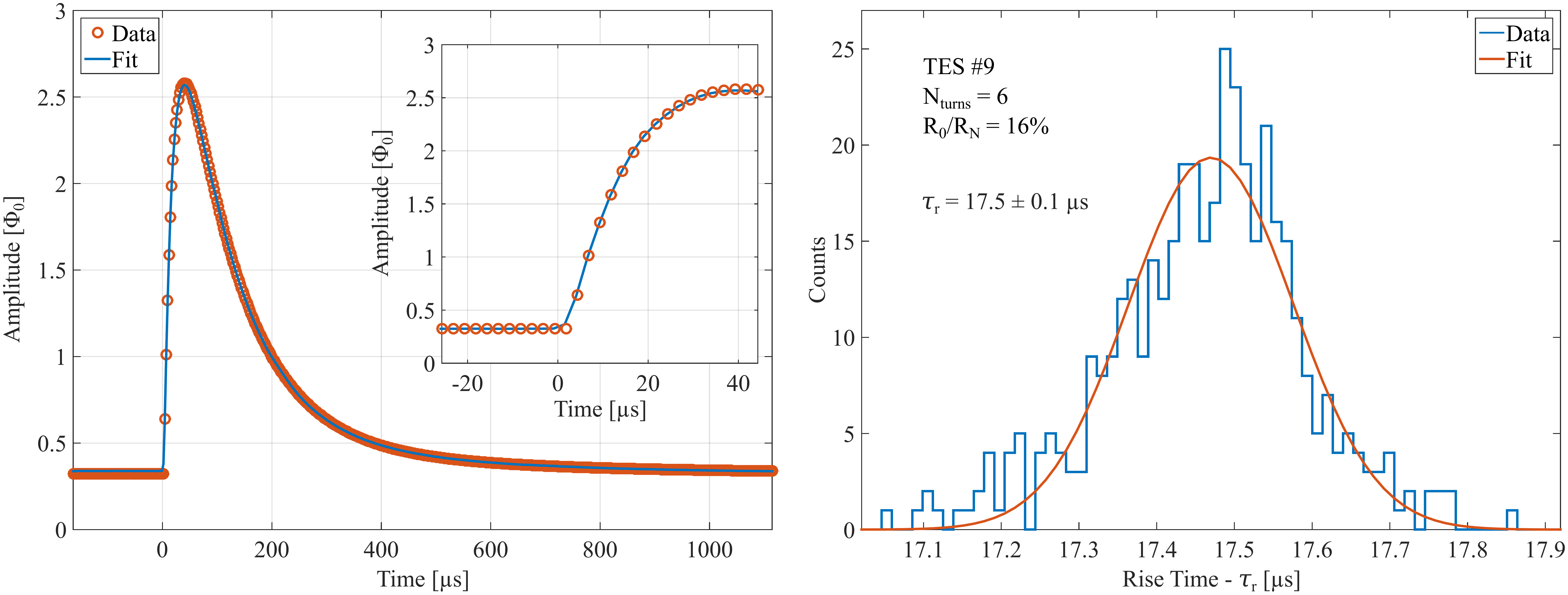}
\caption{\label{fig:timeress} (left) Example of pulse fit with the two-body model (TES \#9). Inlet, Expanded view in the rise region. (right) Rise time distribution with gaussin fit to extrapolate the mean exponential constant (TES \#9).}. 
\end{figure}

As introduced in Section \ref{sec:det}, given a sampling rate of 500\,kHz together with a detector rise and decay times around 10-20\,$\mu$s and 100-150\,$\mu$s, respectively, a time resolution better than 3\,$\mu$s can be achieved, by keeping the pile-up contribution as low as possible. While the TES decay time depends on the pixel geometry and on the thermal design, the rise time depends, at the first order, on the bandwidth of the bias circuit (i.e. $\tau_r\sim R_0/L_N$). Given a working point resistance $R_0$ in the range of 1 to 2\,m$\Omega$, the Nyquist inductor $L_N$ must be properly tuned in order to have a rise time in the range from 10 to 20\,$\mu$s. The choices of $L_N$ are made  selecting a proper number of inductance turns on the IF board. The measured rise and decay constants were obtained by analyzing the pulse shape. Each event from the Mn K$\alpha$ peak was accurately fit by using a two body model~\cite{TwoBody}, that describes the time evolution of the temperature in the TES. With this model, the rise time $\tau_r$ can be expressed as an exponential constant while the decay time by two, namely $\tau_s$ and $\tau_t$. The short decay constant $\tau_s$ is due to the detector thermal recovery, while the long decay constant $\tau_l$ is due to the existence of a weakly coupled subsystem in the absorber. A fit example is reported in figure \ref{fig:timeress} (left); the obtained exponential time constants were then distributed and the mean values were extrapolated by a gaussian fit (figure \ref{fig:timeress}, right). In order to find the best configuration, tests with different values of Nyquist inductor $L_N$ were performed.
Results for the four measured detectors are reported in table \ref{tab:ress}. Measurements were performed with different values of $L_N$ and $R_0/R_N$. As expected, the fastest detector responses were obtained with the lower $L_N$ inductors ($\sim50$\,nH, 6-turn) and higher working point ($20\%R_N$ instead of $15\%R_N$). 

\begin{table}[!t]
\centering
\scalebox{0.85}{  
\begin{tabular}{|c|cc|ccc|cc|cc|} 
\hline
 TES \# & $\Delta E_{\scalebox{.6}{\mbox{FWMH}}}$\,[eV]   & $\Delta E_{\scalebox{.6}{\mbox{FWMH}}}$\,[eV]  & $\tau_r$\,[$\mu$s] & $\tau_s$\,[$\mu$s] & $\tau_l$\,[$\mu$s] & $L_N$\,[nH]  &  $N_{\scalebox{.6}{\mbox{turns}}}$ & $I_{\scalebox{.6}{\mbox{BIAS}}}$\,[m$\Omega$] & $R_0/R_N$\,[\%]\\ 
        & (\textit{expected)}                          & (\textit{measured)}                         &  &  & &  &  & &\\ 
\hline
9   & 4.5 &  $5.2 \pm 0.2$  &  39   & 120  & 548 & 64 & 8 & 498 & 15\% \\
\hline
8   & 5.5 &  $8.4 \pm 0.9$  &  14   & 128  & 580 & 50 & 6 & 607 & 20\%\\
9   & -  &  -   &  18   & 103  & 375 & 50 & 6 & 607 & 20\%\\
11  & 7.6 &  $10.0\pm 3.4$  &  32   & 115  & 738 & 50 & 6 & 522 & 15\%\\
19  & 5.2 &  $9.5 \pm 3.2$  &  32   & 102  & 823 & 50 & 6 & 522 & 15\%\\
\hline
\end{tabular}
}
\caption{\label{tab:ress} Examples of energy and time resolutions at the 5.9\,keV peak for the four acquired TESs. The measured energy resolution is compared to the expected value from the signal-to-noise ratio. (All measurements performed at $T_0=40$\,mK)}
\end{table}

After several improvements on the thermal coupling, on the correction and analysis algorithms, and on the read out system, preliminary results from measurements currently in progress on the same pixels, but read out by using a digital 16-channel multiplexing set-up (Section \ref{sec:1000pixels} for more details) showed better results with an energy resolution close to 4\,eV and a rise time within the range 10-20\,$\mu$s~\cite{HOLMES_EPJC}. The final HOLMES $4\times 16$ sub-arrays will be produced with detectors similar to TES \#9.  

\section{Scaling to 1000 pixels}\label{sec:1000pixels}
The developed 2-channel read out demonstrated that the microwave multiplexing can fulfill the HOLMES requirements in terms of energy and time resolutions and detector noise level. It is based on commercial analog components, and it represented an important step in the demonstratin of the techniques but, at the same time, it was exploited to perform preliminary characterization on the first TES detector arrays specifically designed for HOLMES. Nevertheless this approach is not scalable to 1024 pixels, the final goal for HOLMES. A key enabling technology for reading out large-scale arrays is the microwave multiplexing based on software-defined radio (SDR) techniques. Digital signal processing algorithms on a field-programmable gate array (FPGA) replaces analog components to implement many functions needed to demodulate and reconstruct individual sensor signals, as introduced in the previous sections. High-speed digital-to-analog (ADCs) and analog-to-digital (DACs) converters enable the generation and sampling of probe tones.

\begin{figure}[!t]
\centering\includegraphics[width=\textwidth,clip]{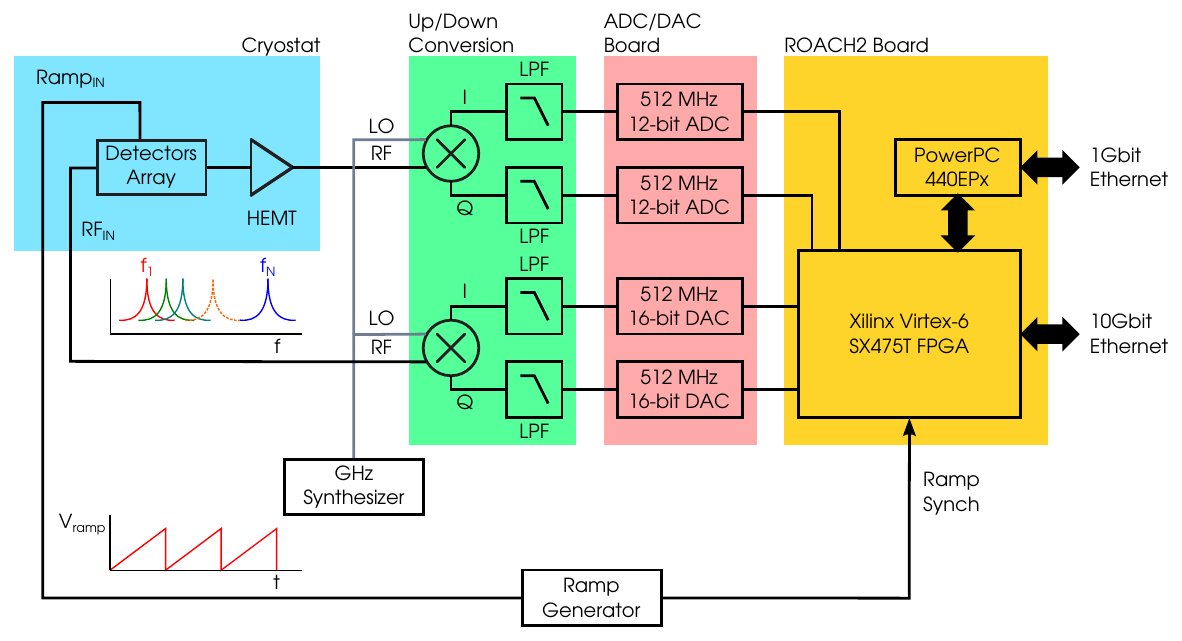}
\caption{\label{fig:roach} Block diagram for a microwave multiplexing readout system based on software-defined radio techniques.}
\end{figure}

The scheme of a SDR-based microwave multiplexing system is shown in figure \ref{fig:roach}. The multiplexing is based on a heterodyne mixing scheme. The sinusoidal probe tones, one for each microresonator to read out, are generated by a FPGA by implementing a Direct Digital Synthesizer (DDS)~\cite{DDS}. DDS is a digital technique that allows to generate a frequency- and phase-tunable output signal referenced to a fixed-frequency precision clock source. The resulting  components are converted in a baseband frequency range signal (e.g., 10-512\,MHz) by two high speed digital-to-analog converter. The up-conversion in the RF frequency range (4-8\,GHz), where the multiplexer chip works, is performed by mixing I and Q with a local oscillator (LO). The tones are then sent (RF\textsubscript{IN}) into the cryostat and through the microwave multiplexer feedline in the detector box. The output modulated tones (RF\textsubscript{OUT}) are mixed down from RF to baseband by exploiting the same LO signal. Two high speed analog-to-digital converters finally digitize the entire baseband signal consisting of all the tones. The FPGA firmware implements signal processing algorithms that reconstruct each sensor response from the modulated tones by extracting the amplitude and phase at each resonator frequency and deconvolving the flux ramp modulation. Separation of the tone frequencies (called channelization) is typically performed by implementing a Digital Down Conversion (DDC)~\cite{DDC} or a Polyphase Filter Bank (PFB)~\cite{POLYPHASE}. 

The HOLMES read out and multiplexing system is based on electronics developed for the readout of microwave kinetic inductance detectors (MKIDs) for the MUSIC~\cite{MUSIC} and ARCONS~\cite{ARCONS} experiments.
These readout systems are themselves based on the 2nd-generation Reconfigurable Open Architecture Computing Hardware (ROACH2) platform developed and maintained by the Collaboration for Astronomy Signal Processing for Electronics Research Consortium (CASPER)~\cite{CASPER}. The system is composed by a digital board (ROACH2), that accommodates a Xilinx Virtex 6 FPGA for the signal processing and a PowerPC 440EPx for the slow control, and by a peripheral DAC/ADC board that hosts two DACs (1000\,MS/s, 16\,bit, 75\,dBc of NSD) and two ADCs (550\,MS/s, 12\,bits, 64\,dB of SNR) to generate and acquire the in-phase and quadrature signals. An intermediate frequency (IF) circuit is used to combine the I and Q components and to mix them with a local oscillator (LO) up (down) to (from) the resonator frequencies. In order to synchronize all the components of the system (ADCs, DACs and FPGA and LOs) a common external clock provided by a rubidium frequency standard is used. The acquired data are transmitted from the ROACH2 board to the data acquisition computer through a 10\,Gb/s ethernet connection based on a hot-pluggable optical module transceiver (Small Form-factor Pluggable, SFP+). 

Due to the limited ADCs bandwidth available ($f_{\scalebox{0.6}{ADC}}=512$\,MHz) and the 14\,MHz resonator spacing needed for HOLMES (to avoid cross-talk between adjacent resonators), only 32 micro-resonators ($\sim1$ multiplexer chip) could be read out per ROACH2. With the given ADC bandwidth and rise time to read out all the 1024 pixels of HOLMES, a total of 32 ROACH2 boards are needed. A similar system developed at NIST has demonstrates the simultaneous multiplexing and read of 128 TESs~\cite{128Mux} (slower than the one developed for HOLMES) without significant degradation of the expected resolution for gamma- and hard X-rays spectroscopy.


A preliminary 16-channel multiplexing system is being set up in Milano. The system uses only half of available ADCs bandwidth in order to match the baseband bandwidth of the amplifiers used after the down-conversions. This setup uses the new $\mu$mux17a multiplexer chip. The development of a 64-channel system based on two ROACH2 boards and on a remotely programmable semi-commercial up- and down-conversion circuitry, designed to fulfill the HOLMES requirements, is also currently in progress. This setup will be fundamental for the read out of the first microcalorimeter $4\times 16$ sub-array, currently in development, with \textsuperscript{163}Ho nuclei implanted.   

\section{Conclusion}
The working principle and the development of a two-channel read out and multiplexing system for low temperature microcalorimeters suitable for the determination of the neutrino mass were introduced. The two-channel system presented was not only used as pure read out demonstrator but also as ready-to-use system to characterize and test the first detectors specifically designed for HOLMES. The obtained results ensured the absence of performance degrading problems due to the read out and confirmed that the proposed read out and multiplexing techniques satisfy the requirements of the HOLMES experiment. The demonstration of a 2-channel read out represented the first important milestone in the HOLMES read out system development. It represented a useful test bench for all the software and algorithms needed for the demodulation processes, setting a baseline for the multichannel FPGA-based digital read out, currently in development,  needed to read out 1024 pixels.

\section*{Acknowledgements}
This work was supported by the European Research Council (FP7/2007-2013) under Grant Agreement HOLMES no. 340321. We also acknowledge the support from INFN through the MARE project and from the NIST Innovations in Measurement Science program for the TES detector development. Mention of commercial products is for information only and does not imply recommendation or endorsement.

\appendix
\section{Flux Ramp Demodulation}\label{sec:appendix-flux-ramp}
For the i\textsuperscript{th}-ramp the demodulated angle variation $\phi_i$ is obtained from the modulated signal $\theta_i(t)$ by using:

\begin{equation}\label{eq:demod}
\phi_i=\arctan{
\left[
- \cfrac
{\sum\limits_{t=1}^{N_{spr}} \theta_i(n) \sin{\left(2\pi\,n_{\Phi_0}f_{r}t\right)}}
{\sum\limits_{t=1}^{N_{spr}} \theta_i(n) \cos{\left(2\pi\,n_{\Phi_0}f_{r}t\right)}}
\right]
}
\end{equation}
\noindent where $N_{spr}$ is the number of sample per ramp, $t$ the t\textsuperscript{th}-sample ($t=1,..,N_{spr}$), and $n_{\Phi_0}f_{r}$ is the number of $\Phi_0$ per ramp multiplied by the ramp frequency, that represents the demodulation carrier frequency. This demodulation formula can be demonstrated as follows. 

Denoting the carrier angular frequency as $\omega_c=2\pi\,n_{\Phi_0}f_{r}$, considering the modulated signal as periodic with the same angular frequency $\theta_i(n)=\cos{\left(\omega_cn+\phi_i\right)}$, and applying the trigonometric sum identity $\cos{\left(\omega_cn+\phi_i\right)}=\cos{\left(\omega_cn\right)}\cos{\phi_i}-\sin{\left(\omega_cn\right)}\sin{\phi_i}$, the \ref{eq:demod} numerator can be written as:

\begin{eqnarray}\label{eq:step1}
\sum\limits_{t=1}^{N_{spr}} \theta_i(t)\sin{\left(\omega_ct\right)}&=&
\sum\limits_{t=1}^{N_{spr}} \cos{\left(\omega_ct+\phi_i\right)}\sin{\left(\omega_ct\right)}\nonumber\\
&=&\cos{\phi_i}\sum\limits_{t=1}^{N_{spr}}\sin{\left(\omega_ct\right)}\cos{\left(\omega_ct\right)}-
\sin{\phi_i}\sum\limits_{t=1}^{N_{spr}}\sin^{2}{\left(\omega_ct\right)}
\end{eqnarray}

\noindent Applying the Werner formulas it is possible to obtain:
\begin{eqnarray}
\sin{\left(\omega_ct\right)}\cos{\left(\omega_ct\right)}&=&\cfrac{1}{2}\left[\sin{2\omega_ct}-\sin{\left(0\right)}\right]=\cfrac{1}{2}\sin{(2\omega_ct)}\label{eq:w1}\\
\sin^{2}{\left(\omega_ct\right)}&=&\cfrac{1}{2}\left[\cos{\left(0\right)}-\cos(2\omega_ct)\right]=\cfrac{1}{2}\left[1-\cos{(2\omega_ct)}\right]\label{eq:w2}
\end{eqnarray}

\noindent Replacing \ref{eq:w1} and \ref{eq:w2} in \ref{eq:step1} the relationship becomes:

\begin{eqnarray}\label{eq:nom}
\sum\limits_{t=1}^{N_{spr}} \theta_i(t)\sin{\left(\omega_ct\right)}&=&
\cfrac{1}{2}\cos{\phi_i}\underbrace{\sum\limits_{t=1}^{N_{spr}}\sin{(2\omega_ct)}}_\text{=0}
-\cfrac{N_{spr}}{2}\sin{\phi_i} 
+\cfrac{1}{2}\sin{\phi_i}\underbrace{\sum\limits_{t=1}^{N_{spr}}\cos{(2\omega_ct)}}_\text{=0}\nonumber\\
&=&-\cfrac{N_{spr}}{2}\sin{\phi_i} 
\end{eqnarray}

\noindent The two sums are null over a ramp period, since within each of them there are an integer number $k$ of carrier oscillations. With similar computations the \ref{eq:dem} denominator can be written as 

\begin{eqnarray}\label{eq:dem}
\sum\limits_{t=1}^{N_{spr}} \theta_i(t)\cos{\left(\omega_ct\right)}=\cfrac{N_{spr}}{2}\cos{\phi_i} 
\end{eqnarray}

\noindent The negative ratio of \ref{eq:nom} and \ref{eq:dem} gives:

\begin{equation}\label{eq:end}
- \cfrac{\sum\limits_{t=1}^{N_{spr}} \theta_i(t)\sin{\left(\omega_ct\right)}}{\sum\limits_{t=1}^{N_{spr}} \theta_i(t)\cos{\left(\omega_ct\right)}}
=-\cfrac{-\cfrac{N_{spr}}{2}\sin{\phi_i}}{\cfrac{N_{spr}}{2}\cos{\phi_i}}=\tan{\phi_i}
\end{equation}

\noindent That demonstrates the demodulation formula \ref{eq:demod}. The flux variation within the i\textsuperscript{th}-ramp period can be computed by reverting the relationship \ref{eq:phi-angle}:

\begin{equation}\label{eq:endend}
\Phi_i=\cfrac{\Phi_0}{2\pi}\,\phi_i
\end{equation}

\noindent Finally, the current variation across the TES can be computed considering the SQUID the transduction gain $M$:

\begin{equation}\label{eq:veryend}
I_{\scalebox{.6}{\mbox{TES}}}(i)=\cfrac{\Phi_i}{M}
\end{equation}

\section{Bandwidth Budget}\label{sec:appendix-budget}
The bandwidth budget is the total bandwidth required to multiplex a number detectors. For the microwave multiplexing it is computed as following. Denoting as

\begin{multicols}{2}
  \begin{itemize}\itemsep0em 
  \item[] $f_s$: sampling rate;                         
  \item[] $f_{\scalebox{.6}{\mbox{ramp}}}$: flux ramp frequency;                  
  \item[] $\Delta f_{\scalebox{.6}{\mbox{BW}}}$: resonator bandwidth;             
  \item[] $n_{\Phi_0}$: number of $\Phi_0$ per ramp;
  \item[] $S$: frequency spacing between tones;        
  \item[] $g_f$: guard factor between tones;                                 
  \item[] $\tau_r$: detector rise time;                                      
  \item[] $R_d$: distortion suppression factor;         
  \item[] $f_{\scalebox{.6}{\mbox{ADC}}}$: ADC bandwidth;                 
  \item[] $n_{\scalebox{.6}{\mbox{TES}}}$: number of detector per ADC board;
  \end{itemize}
\end{multicols}
        
\begin{itemize}
\item The flux ramp sets the sampling frequency: $f_s = f_{\scalebox{.6}{\mbox{ramp}}}$;
\item The resonator bandwidth and the number of SQUID oscillations within a ramp period must satisfy the Nyquist-Shannon sampling theorem:
\begin{equation}\label{eq:shannon}
\Delta f_{\scalebox{.6}{\mbox{BW}}}\ge 2 f_{\scalebox{.6}{\mbox{ramp}}}\,n_{\Phi_0}
\end{equation}
\item To avoid cross-talk the spacing between frequency adjacent resonances must be grater than the resonator bandwidth. Denoting as $g_f$ the guard factor, the spacing results as 
\begin{equation}\label{eq:space}
S>g_f\Delta f_{\scalebox{.6}{\mbox{BW}}}
\end{equation}
\item The sampling time must be slower than the rise time to avoid distortion. Denoting as $R_d$ the distortion suppression factor (2 is Nyquist-Shannon limit), the sampling frequency can be written as
\begin{equation}\label{eq:samp}
f_s=f_{\scalebox{.6}{\mbox{ramp}}}\textbf{}\ge\cfrac{R_d}{\tau_r}  
\end{equation}
\item Rewriting the \ref{eq:space} by replacing \ref{eq:shannon} and \ref{eq:samp} one obtains
\begin{equation}\label{eq:space1}
S=\cfrac{2 n_{\Phi_0}g_f R_d}{\tau_r}
\end{equation}
\item The number of multiplexable TES detectors for an ADC with bandwidth $f_{\scalebox{.6}{\mbox{ADC}}}$ and for a resonance frequency spacing $S$ is given by
\begin{equation}\label{eq:npixel}
n_{\scalebox{.6}{\mbox{TES}}}=\cfrac{f_{\scalebox{.6}{\mbox{ADC}}}}{S}
\end{equation}
\item Finally, replacing \ref{eq:space1} in \ref{eq:npixel}, the multiplexing factor per ADC board can be written as
\begin{equation}\label{eq:npixel_final}
n_{\scalebox{.6}{\mbox{TES}}}=\cfrac{f_{\scalebox{.6}{\mbox{ADC}}}\cdot\tau_r}{2\cdot n_{\Phi_0}\cdot g_f\cdot R_d}
\end{equation}

\end{itemize}

\section{Resonance Fit Model}\label{sec:appendix-res-model}
 Resonance parameters as the resonance frequency $f_{\scalebox{.6}{\mbox{res}}}$, the internal quality factor $Q_i$ and the coupling quality factor $Q_c$ can be estimated by fitting the measured forward transmission $S_{21}$. Petersan \textit{et al.}~\cite{CIRCLE_FIT} provide a quantitative comparison of the different possible methods to perform the fit. Among them, in this work a modified version of the fit procedure proposed in~\cite{GAO_THESIS} is used. In this approach the transmission $S_{21}$ through the microresonator, amplifiers, and cables measured performing a frequency scan, can be written as: 

\begin{equation}\label{eq:fit-model}
  \begin{split}
    \left|S_{21}\right|(f)& =\left|I+\imath Q\right|(f)=\\
                         & =af+bf^2+cf^3+A\cdot\left[
      1-Q_t\left(\cfrac{1}{Q_t}-\cfrac{1}{Q_i}\right)
      \cdot\cfrac{e^{\imath\varphi_0}}{1-2\imath\,Q_t\left(\cfrac{f-f_{\scalebox{.6}{\mbox{res}}}}{f_{\scalebox{.6}{\mbox{res}}}}\right)+\imath L} 
      \right]
  \end{split}
\end{equation}

\noindent where $a,b,c$ are the coefficients of the third order polynomial correction to the $S_{21}$ off-resonance, $A$ is a normalization factor, $Q_t$ and $Q_i$ are the total and internal quality factor, respectively ($Q_t=(Q_c^{-1}+Q_i^{-1})^{-1}$, where $Q_c$ is the coupling quality factor), $\varphi_0$ is a rotation angle due to impedance mismatches, $f_{\scalebox{.6}{\mbox{res}}}$ is the resonance frequency, and $L$ a parasitic impedance that takes into account a skewing due to the presence of the resonator itself, which modifies both the electric and magnetic fields around the feedline, acting as a lumped complex impedance on the line~\cite{GIORDANO_THESIS}. The resonator bandwidth is then computed ad $\Delta f_{\scalebox{.6}{\mbox{BW}}} =f_{\scalebox{.6}{\mbox{res}}}/Q_t$. The $f_{\scalebox{.6}{\mbox{res}}}$ value is first extrapolated with a second order fit in narrow range across the resonance minimum and then fixed in the final fit procedure.

\bibliographystyle{JHEP}
\bibliography{umux}

\providecommand{\href}[2]{#2}\begingroup\raggedright\begin{thebibliography}{10}

\bibitem{Capozzi2017}
F.~Capozzi et~al., {\it {Global constraints on absolute neutrino masses and
  their ordering}},  {\em Phys. Rev. D} {\bf 95} (2017), no.~9 096014,
  [\href{http://xxx.lanl.gov/abs/1703.04471}{{\tt arXiv:1703.04471}}].

\bibitem{deSalas2017}
P.~F. de~Salas et~al., {\it {Status of neutrino oscillations 2018: 3$\sigma$
  hint for normal mass ordering and improved CP sensitivity}},  {\em Phys.
  Lett. B} {\bf 782} (2018) 633--640,
  [\href{http://xxx.lanl.gov/abs/1708.01186}{{\tt arXiv:1708.01186}}].

\bibitem{Capozzi2018}
F.~Capozzi and others., {\it {Current unknowns in the three neutrino
  framework}},  {\em Prog. Part. Nucl. Phys.} {\bf 102} (2018) 48--72,
  [\href{http://xxx.lanl.gov/abs/1804.09678}{{\tt arXiv:1804.09678}}].

\bibitem{Weinheimer2013}
C.~Weinheimer and K.~Zuber, {\it {Neutrino Masses}},  {\em Annalen Phys.} {\bf
  525} (2013), no.~8-9 565--575, [\href{http://xxx.lanl.gov/abs/1307.3518}{{\tt
  arXiv:1307.3518}}].

\bibitem{Drexlin2013}
G.~Drexlin et~al., {\it {Current direct neutrino mass experiments}},  {\em Adv.
  High Energy Phys.} {\bf 2013} (2013) 293986,
  [\href{http://xxx.lanl.gov/abs/1307.0101}{{\tt arXiv:1307.0101}}].

\bibitem{Mainz}
C.~Kraus et~al., {\it {Final results from phase II of the Mainz neutrino mass
  searchin tritium $\beta$ decay}},  {\em Eur. Phys. J. C} {\bf 40} (2005),
  no.~4 447--468.

\bibitem{Mainz2}
C.~Kraus, A.~Singer, K.~Valerius, and C.~Weinheimer, {\it {Limit on sterile
  neutrino contribution from the Mainz Neutrino Mass Experiment}},  {\em Eur.
  Phys. J. C} {\bf 73} (2013) 2323,
  [\href{http://xxx.lanl.gov/abs/1210.4194}{{\tt arXiv:1210.4194}}].

\bibitem{Troitsk}
V.~Lobashev et~al., {\it Direct search for neutrino mass and anomaly in the
  tritium beta-spectrum: Status of troitsk neutrino mass experiment},  {\em
  Nucl. Phys. B. (Proc. Suppl.)} {\bf 91} (2001), no.~1-3 280--286.

\bibitem{Troitsk2}
 Troitsk Collaboration Collaboration, V.~Aseev et~al., {\it {An upper limit on
  electron antineutrino mass from Troitsk experiment}},  {\em Phys.Rev. D} {\bf
  84} (2011) 112003, [\href{http://xxx.lanl.gov/abs/1108.5034}{{\tt
  arXiv:1108.5034}}].

\bibitem{KATRIN}
J.~Angrik et~al., {\it {KATRIN design report 2004}},  {\em Tech. Rep.,
  Forschungszentrum, Karlsruhe, Germany} (2004) 447--468. {NPI ASCR
  \v{R}e\v{z}, EXP-01/2005, FZKA Scientific Report 7090, MS-KP-0501}.

\bibitem{KATRIN2017}
L.~Bornschein et~al., {\it {Status of the Karlsruhe Tritium Neutrino Mass
  Experiment KATRIN}},  {\em Fusion Science and Technology} {\bf 71} (2017),
  no.~4 485--490.

\bibitem{Project8}
B.~Monreal and J.~A. Formaggio, {\it {Relativistic Cyclotron Radiation
  Detection of Tritium Decay Electrons as a New Technique for Measuring the
  Neutrino Mass}},  {\em Phys. Rev. D} {\bf 80} (2009) 051301,
  [\href{http://xxx.lanl.gov/abs/0904.2860}{{\tt arXiv:0904.2860}}].

\bibitem{Project8_3}
 Project 8 Collaboration, A.~Ashtari~Esfahani et~al., {\it {Determining the
  neutrino mass with cyclotron radiation emission spectroscopy-Project 8}},
  {\em J. Phys. G} {\bf 44} (2017), no.~5 054004,
  [\href{http://xxx.lanl.gov/abs/1703.02037}{{\tt arXiv:1703.02037}}].

\bibitem{Project8_2}
 Project 8 Collaboration, D.~Asner et~al., {\it {Single electron detection and
  spectroscopy via relativistic cyclotron radiation}},  {\em Phys. Rev. Lett.}
  {\bf 114} (2015), no.~16 162501,
  [\href{http://xxx.lanl.gov/abs/1408.5362}{{\tt arXiv:1408.5362}}].

\bibitem{UseOfLTD}
A.~Nucciotti, {\it The use of low temperature detectors for direct measurements
  of the mass of the electron neutrino},  {\em Advances in High Energy Physics}
  {\bf 2016} (2016) 9153024, [\href{http://xxx.lanl.gov/abs/1511.00968}{{\tt
  arXiv:1511.00968}}].

\bibitem{HoProposal}
A.~D. R\'{u}jula and M.~Lusignoli, {\it {Calorimetric measurements of
  163holmium decay as tools to determine the electron neutrino mass}},  {\em
  Physics Letters B} {\bf 118} (1982), no.~4-6 429--434.

\bibitem{HOLMES}
 HOLMES Collaboration, B.~Alpert et~al., {\it {HOLMES - The Electron Capture
  Decay of $^{163}$Ho to Measure the Electron Neutrino Mass with sub-eV
  sensitivity}},  {\em The European Physical Journal C} {\bf 75} (2015), no.~3
  112, [\href{http://xxx.lanl.gov/abs/1412.5060}{{\tt arXiv:1412.5060}}].

\bibitem{ECHo}
L.~Gastaldo et~al., {\it The electron capture in 163ho experiment -- echo},
  {\em Eur. Phys. J. Special Topics} {\bf 226} (Jun, 2017) 1623--1694.

\bibitem{MMC_REVIEW}
A.~Fleischmann, C.~Enss, and G.~Seidel, {\it Metallic magnetic calorimeters},
  {\em Topics in Applied Physics} {\bf 99} (2005) 151--216.

\bibitem{HOLMES_Ho}
S.~Heinitz et~al., {\it {Production and separation of 163Ho for nuclear physics
  experiments}},  {\em PLoS One} {\bf 13} (2018), no.~8 e0200910.

\bibitem{Engle2013}
J.~W. Engle et~al., {\it {Evaluation of $^{163}$Ho production options for
  neutrino mass measurements with microcalorimeter detectors}},  {\em Nucl.
  Instrum. Meth. B} {\bf 311} (2013) 131--138.

\bibitem{NuMECS}
M.~P. Croce et~al., {\it {Development of holmium-163 electron-capture
  spectroscopy with transition-edge sensors}},  {\em Journal of Low Temperature
  Physics} (2016) 1--11, [\href{http://xxx.lanl.gov/abs/1510.03874}{{\tt
  arXiv:1510.03874}}].

\bibitem{Nucciotti_sens}
A.~Nucciotti, {\it {Statistical sensitivity of $^{163}$Ho electron capture
  neutrino mass experiments}},  {\em Eur. Phys. J. C} {\bf 74} (2014), no.~11
  3161, [\href{http://xxx.lanl.gov/abs/1405.5060}{{\tt arXiv:1405.5060}}].

\bibitem{TES_MODEL}
K.~Irwin and G.~Hilton, {\it Transition-edge sensors},  {\em Topics in Applied
  Physics} {\bf 99} (2005) 63--150.

\bibitem{TES_REVIEW}
J.~N. Ullom and D.~A. Bennett, {\it Review of superconducting transition-edge
  sensors for x-ray and gamma-ray spectroscopy},  {\em Superconductor Science
  and Technology} {\bf 28} (2015), no.~8 084003.

\bibitem{ETF}
K.~D. Irwin, {\it An application of electrothermal feedback for high resolution
  cryogenic particle detection},  {\em Applied Physics Letters} {\bf 66}
  (1995), no.~15 1998--2000.

\bibitem{SQUID}
J.~Clarke and A.~I. Braginski, {\em The SQUID handbook, Volume 1}.
\newblock Wiley-Vch, 1st edition~ed., 2006.

\bibitem{DCSQUID}
R.~C. Jaklevic, J.~Lambe, A.~H. Silver, and J.~E. Mercereau, {\it {Quantum
  Interference Effects in Josephson Tunneling}},  {\em Physical Review Letters}
  {\bf 12} (1964) 159--160.

\bibitem{RFSQUID}
J.~E. Zimmerman and A.~H. Silver, {\it {Coherence and Quantization in Nearly
  Superconducting Rings}},  {\em Physical Review} {\bf 167} (1968) 418--420.

\bibitem{MATES_THESIS}
J.~A.~B. Mates, {\em {The Microwave SQUID Multiplexer}}.
\newblock PhD thesis, {Department of Physics, University of Colorado}, Libby
  Dr, Boulder, CO 80302, 2011.

\bibitem{MSQUID}
J.~A.~B. Mates, G.~C. Hilton, K.~D. Irwin, L.~R. Vale, and K.~W. Lehnert, {\it
  Demonstration of a multiplexer of dissipationless superconducting quantum
  interference devices},  {\em Applied Physics Letters} {\bf 92} (2008), no.~2
  023514.

\bibitem{Shannon}
K.~D. Irwin, {\it Shannon limits for low-temperature detector readout},  {\em
  AIP Conference Proceedings} {\bf 1185} (2009), no.~1 229--236.

\bibitem{TDM}
J.~A. Chervenak, K.~D. Irwin, E.~N. Grossman, J.~M. Martinis, C.~D. Reintsema,
  and M.~E. Huber, {\it Superconducting multiplexer for arrays of transition
  edge sensors},  {\em Applied Physics Letters} {\bf 74} (1999), no.~26
  4043--4045.

\bibitem{FDM}
J.~Yoon et~al., {\it Single superconducting quantum interference device
  multiplexer for arrays of low-temperature sensors},  {\em Applied Physics
  Letters} {\bf 78} (2001), no.~3 371--373.

\bibitem{CDM}
M.~D. Niemack et~al., {\it {Code-division SQUID multiplexing}},  {\em Applied
  Physics Letters} {\bf 96} (2010), no.~16 163509.

\bibitem{Doriese2016}
W.~B. Doriese et~al., {\it {Developments in Time-Division Multiplexing of X-ray
  Transition-Edge Sensors}},  {\em Journal of Low Temperature Physics} {\bf
  184} (2016), no.~1 389--395.

\bibitem{beam}
W.~B. Doriese et~al., {\it {A practical superconducting-microcalorimeter X-ray
  spectrometer for beamline and laboratory science}},  {\em Review of
  Scientific Instruments} {\bf 88} (2017), no.~5 053108.

\bibitem{ATHENA_MUX}
J.~van~der Kuur et~al., {\it {Optimising the multiplex factor of the frequency
  domain multiplexed readout of the TES-based microcalorimeter imaging array
  for the X-IFU instrument on the Athena x-ray observatory}},
  \href{http://xxx.lanl.gov/abs/1611.05268}{{\tt arXiv:1611.05268}}.

\bibitem{ATHENA}
D.~Barret et~al., {\it The athena x-ray integral field unit (x-ifu)},  {\em
  Proc. SPIE} {\bf 9905} (2016).

\bibitem{IrwinMSQUID}
K.~D. Irwin and K.~W. Lehnert, {\it Microwave squid multiplexer},  {\em Applied
  Physics Letters} {\bf 85} (2004), no.~11 2107--2109.

\bibitem{HahnMSQUID}
I.~Hahn, P.~Day, B.~Bumble, and H.~G. LeDuc, {\it Recent results of a new
  microwave squid multiplexer},  {\em Journal of Low Temperature Physics} {\bf
  151} (2008), no.~3 934--939.

\bibitem{DayMKIDS}
P.~K. Day, H.~G. LeDuc, B.~A. Mazin, A.~Vayonakis, and J.~Zmuidzinas, {\it A
  broadband superconducting detector suitable for use in large arrays},  {\em
  Nature} {\bf 425} (2003), no.~6960 817--821.

\bibitem{ZmuidzinasMKIDS}
J.~Zmuidzinas, {\it Superconducting microresonators: Physics and applications},
   {\em Annual Review of Condensed Matter Physics} {\bf 3} (2012) 169--214.

\bibitem{RAMP}
J.~Mates et~al., {\it Flux-ramp modulation for squid multiplexing},  {\em
  Journal of Low Temperature Physics} {\bf 167} (2012), no.~5-6 707--712.

\bibitem{RAMP2}
K.~W. Lehnert, K.~D. Irwin, M.~A. Castellanos-Beltran, J.~A.~B. Mates, and
  L.~R. Vale, {\it Evaluation of a microwave squid multiplexer prototype},
  {\em IEEE Transactions on Applied Superconductivity} {\bf 17} (2007), no.~2
  705--709.

\bibitem{Pozar}
D.~M. Pozar, {\em Microwave Engineering}.
\newblock John Wiley \& Sons Inc, 3rd edition~ed., 2009.

\bibitem{MMC}
S.~Kempf, M.~Wegner, L.~Gastaldo, A.~Fleischmann, and C.~Enss, {\it
  {Multiplexed readout of MMC detector arrays using non-hysteretic rf-SQUIDs}},
   {\em Journal of Low Temperature Physics} {\bf 176} (2014) 426,
  [\href{http://xxx.lanl.gov/abs/1309.4929}{{\tt arXiv:1309.4929}}].

\bibitem{TLS1}
J.~Gao et~al., {\it {Noise properties of superconducting coplanar waveguide
  microwave resonators}},  {\em Applied Physics Letters} {\bf 90} (2007),
  no.~10 102507.

\bibitem{TLS2}
J.~Burnett and other, {\it {Evidence for interacting two-level systems from the
  1/f noise of a superconducting resonators}},  {\em Nature Communications}
  {\bf 4} (2014) 4119.

\bibitem{TLS_EMPIRICAL}
J.~Gao et~al., {\it {A semiempirical model for two-level system noise in
  superconducting microresonators}},  {\em Appl. Phys. Lett} {\bf 92} (2008),
  no.~21 212504.

\bibitem{TLS4}
J.~Gao and other, {\it {Experimental evidence for a surface distribution of
  two-level systems in superconducting lithographed microwave resonators}},
  {\em Appl. Phys. Lett} {\bf 92} (2008), no.~15 152505.

\bibitem{TLS_REDUCTION}
O.~Noroozian et~al., {\it {Two-level system noise reduction for Microwave
  Kinetic Inductance Detectors}},  {\em AIP Conf. Proc} {\bf 1185} (2009),
  no.~1 148--151.

\bibitem{Zmuidzinas2012}
J.~Zmuidzinas, {\it {Superconducting Microresonators: Physics and
  Applications}},  {\em Annu. Rev. Condens. Matter Phys} {\bf 3} (2012), no.~1
  169--214.

\bibitem{PUIU_THESIS}
A.~P. Puiu, {\em {Transition Edge Sensor Calorimeters for HOLMES}}.
\newblock PhD thesis, {Department of Physics, University of Milano-Bicocca},
  Piazza della Scienza 3, I-20126 Milano, 2017.

\bibitem{128Mux}
J.~A.~B. Mates et~al., {\it {Simultaneous readout of 128 X-ray and gamma-ray
  transition-edge microcalorimeters using microwave SQUID multiplexing}},  {\em
  Applied Physics Letters} {\bf 111} (2017), no.~6 062601.

\bibitem{GAMMA-DEM}
O.~Noroozian et~al., {\it {High-resolution gamma-ray spectroscopy with a
  microwave-multiplexed transition-edge sensor array}},  {\em Applied Physics
  Letters} {\bf 103} (2013) 202602,
  [\href{http://xxx.lanl.gov/abs/1310.7287}{{\tt arXiv:1310.7287}}].

\bibitem{Argonne}
T.~J. Madden et~al., {\it {Development of ROACH Firmware for Microwave
  Multiplexed X-Ray TES Microcalorimeters}},  {\em IEEE Trans. Appl. Supercond}
  {\bf 27} (2017), no.~4 1--4.

\bibitem{LosAlamos}
P.~Guss et~al., {\it {High-resolution photon spectroscopy with a
  microwave-multiplexed 4-pixel transition edge sensor array}},  {\em Proc.
  SPIE} {\bf 10393} (2017) 10393--10393--16.

\bibitem{ARCONS}
S.~McHugh et~al., {\it {A readout for large arrays of microwave kinetic
  inductance detectors}},  {\em Review of Scientific Instruments} {\bf 83}
  (2012), no.~4 044702, [\href{http://xxx.lanl.gov/abs/1203.5861}{{\tt
  arXiv:1203.5861}}].

\bibitem{LNF_LNC4_8C}
Low Noise Factory, {\em {LNF-LNC4\_8C, 4-8 GHz Cryogenic Low Noise Amplifier}}.
\newblock Rev. March 2017.

\bibitem{MATES_SEMINAR}
J.~A.~B. Mates, {\it {Microwave SQUID Multiplexing for the HOLMES Experiment}},
   June, 2016.
\newblock {Invited seminar at the INFN Milano-Bicocca}.

\bibitem{McCammon1984}
S.~H. Moseley, J.~C. Mather, and D.~McCammon, {\it Thermal detectors as x-ray
  spectrometers},  {\em J. Appl. Phys.} {\bf 56} (1984), no.~5 1257--1262.

\bibitem{Giachero2017}
A.~Giachero, {\it {Assess the neutrino mass with micro and macro calorimeter
  approach}},  {\em J. Phys. Conf. Ser.} {\bf 841} (2017), no.~1 012027,
  [\href{http://xxx.lanl.gov/abs/1703.02747}{{\tt arXiv:1703.02747}}].

\bibitem{HOLMES_EPJC}
B.~Alpert et~al., {\it {High-resolution high-speed microwave-multiplexed low
  temperature microcalorimeters for the HOLMES experiment}},  {\em Eur. Phys.
  J. C} {\bf 79} (2019), no.~4 304.

\bibitem{Bennett2012}
D.~A. Bennett, D.~S. Swetz, R.~D. Horansky, D.~R. Schmidt, and J.~N. Ullom,
  {\it {A Two-Fluid Model for the Transition Shape in Transition-Edge
  Sensors}},  {\em Journal of Low Temperature Physics} {\bf 167} (May, 2012)
  102--107.

\bibitem{Hays-Wehle2016}
J.~P. Hays-Wehle, D.~R. Schmidt, J.~N. Ullom, and D.~S. Swetz, {\it {Thermal
  Conductance Engineering for High-Speed TES Microcalorimeters}},  {\em J. Low.
  Temp. Phys.} {\bf 184} (2016), no.~1-2 492--497.

\bibitem{Giachero2016}
 HOLMES Collaboration, A.~Giachero et~al., {\it {Measuring the electron
  neutrino mass with improved sensitivity: the HOLMES experiment}},  {\em
  JINST} {\bf 12} (2017), no.~02 C02046,
  [\href{http://xxx.lanl.gov/abs/1612.03947}{{\tt arXiv:1612.03947}}].

\bibitem{Puiu2017}
A.~Puiu et~al., {\it {Development of transition edge sensors with rf-SQUID
  based multiplexing system for the HOLMES experiment}},  {\em J. Phys. Conf.
  Ser.} {\bf 888} (2017), no.~1 012069.

\bibitem{Lindeman2004}
M.~A. Lindeman and othres, {\it Impedance measurements and modeling of a
  transition-edge-sensor calorimeter},  {\em Review of Scientific Instruments}
  {\bf 75} (2004), no.~5 1283--1289.

\bibitem{unexplained}
J.~N. Ullom and othres, {\it Characterization and reduction of unexplained
  noise in superconducting transition-edge sensors},  {\em Applied Physics
  Letters} {\bf 84} (2004), no.~21 4206--4208.

\bibitem{Alpert2015}
B.~Alpert et~al., {\it {Algorithms for Identification of Nearly-Coincident
  Events in Calorimetric Sensors}},  {\em J. Low. Temp. Phys.} {\bf 184}
  (2016), no.~1-2 263--273, [\href{http://xxx.lanl.gov/abs/1512.01608}{{\tt
  arXiv:1512.01608}}].

\bibitem{Ferri2016}
E.~Ferri et~al., {\it {Pile-Up Discrimination Algorithms for the HOLMES
  Experiment}},  {\em J. Low. Temp. Phys.} {\bf 184} (2016), no.~1-2 405--411.

\bibitem{Transformer}
P.~Carniti et~al., {\it {Transformer coupling and its modelling for the
  flux-ramp modulation of rf-SQUIDs}},  {\em Instruments} {\bf 3} (2018), no.~1
  3, [\href{http://xxx.lanl.gov/abs/1712.01577}{{\tt arXiv:1712.01577}}].

\bibitem{linear}
G.~Pessina, {\it Low-noise, low drift, high precision linear bipolar ($\pm$10
  v) voltage supply/reference for cryogenic front-end apparatus},  {\em Review
  of Scientific Instruments} {\bf 70} (1999), no.~8 3473--3478.

\bibitem{Kasa}
I.~K\r{a}sa, {\it A circle fitting procedure and its error analysis},  {\em
  IEEE Trans. Instrum. Meas} {\bf IM-25} (1976), no.~1 8--14.

\bibitem{Taubin}
G.~Taubin, {\it Estimation of planar curves, surfaces, and nonplanar space
  curves defined by implicit equations with applications to edge and range
  image segmentation},  {\em IEEE Trans. Pattern Anal. Mach. Intell.} {\bf 13}
  (1991), no.~11 1115--1138.

\bibitem{Gatti_Manfredi}
E.~Gatti and P.~F. Manfredi, {\it Processing the signals from solid-state
  detectors in elementary-particle physics},  {\em La Rivista del Nuovo
  Cimento} {\bf 9} (1986), no.~1 1--146.

\bibitem{peakfit}
E.~Ferri et~al., {\it {Investigation of peak shapes in the MIBETA experiment
  calibrations}},  {\em The European Physical Journal A} {\bf 48} (2012),
  no.~10 131, [\href{http://xxx.lanl.gov/abs/1210.5875}{{\tt
  arXiv:1210.5875}}].

\bibitem{TwoBody}
D.~A. Bennett et~al., {\it An analytical model for pulse shape and
  electrothermal stability in two-body transition-edge sensor
  microcalorimeters},  {\em Applied Physics Letters} {\bf 97} (2010), no.~10
  102504.

\bibitem{DDS}
K.~A.~H. Jouko~Vankka, {\em Direct Digital Synthesizers: Theory, Design and
  Applications}.
\newblock Springer US, 1st edition~ed., 2001.

\bibitem{DDC}
S.~K. Mitra, {\em Digital Signal Processing: A Computer Based Approach}.
\newblock McGraw-Hill, 4th edition~ed., 2011.

\bibitem{POLYPHASE}
K.~Ad\'{a}mek, J.~Novotn\'{y}, and W.~Armour, {\it A polyphase filter for
  many-core architectures},  {\em Astronomy and Computing} {\bf 16} (2016)
  1--16.

\bibitem{MUSIC}
R.~Duan et~al., {\it An open-source readout for mkids},  {\em Proc. SPIE} {\bf
  7741} (2010) 7741--7741--10.

\bibitem{CASPER}
J.~Hickish et~al., {\it {A Decade of Developing Radio-Astronomy Instrumentation
  using CASPER Open-Source Technology}},  {\em Journal of Astronomical
  Instrumentation} {\bf 05} (2016), no.~04 1641001,
  [\href{http://xxx.lanl.gov/abs/1611.01826}{{\tt arXiv:1611.01826}}].

\bibitem{CIRCLE_FIT}
P.~J. Petersan and S.~M. Anlage, {\it Measurement of resonant frequency and
  quality factor of microwave resonators: Comparison of methods},  {\em Journal
  of Applied Physics} {\bf 84} (September, 1998) 3392--3402.

\bibitem{GAO_THESIS}
J.~Gao, {\em The physics of superconducting microwave resonators}.
\newblock PhD thesis, California Institute of Technology, Pasadena, CA 91125,
  United States, June, 2008.

\bibitem{GIORDANO_THESIS}
C.~Giordano, {\em MKID arrays: panoramic detectors for CMB experiments}.
\newblock PhD thesis, Sapienza University of Rome, Piazzale Aldo Moro, 5, 00185
  Roma, Italy, June, 2009.

\end{thebibliography}\endgroup

\end{document}